\documentclass[journal,draftcls,onecolumn]{IEEEtran}

\usepackage{cite}       

\ifx\pdfoutput\undefined
\usepackage{graphicx}
\else
\usepackage[pdftex]{graphicx}
\fi

\usepackage{psfrag}     

\usepackage{subfigure}  

\usepackage{url}        
\usepackage{amsmath}    
\usepackage{array}

\usepackage{epsfig}
\usepackage{amsfonts}
\usepackage{amssymb}
\usepackage{rotating}
\usepackage{multirow}
\usepackage{bigstrut}
\usepackage{bigdelim}
\usepackage{color}
\usepackage{threeparttable}
\usepackage{footnote}

\newtheorem{example}{Example}
\newtheorem{lemma}{Lemma}

\newtheorem{definition}{Definition}
\newtheorem{theorem}{Theorem}

\newtheorem{construction}{Construction}

\DeclareMathOperator{\rank}{rank}
\DeclareMathOperator{\diag}{diag}

\newcommand\T{\rule{0pt}{2.6ex}}
\newcommand\B{\rule[-1.2ex]{0pt}{0pt}}

\begin{document}
\title{Turbo Decoding on the Binary Erasure Channel: Finite-Length Analysis and Turbo Stopping Sets}
\author{Eirik~Rosnes,~\IEEEmembership{Member,~IEEE,}
        and~{\O}yvind~Ytrehus,~\IEEEmembership{Senior Member,~IEEE}%
\thanks{Parts of this work have been presented at the 2005 IEEE International Symposium on Information Theory, Adelaide, SA, Australia, September 2005.} 
\thanks{This work was carried out within NEWCOM and supported by Nera Networks AS and the Norwegian Research Council (NFR) Grants 156712/220, 164803/V30, and 146874/420.}%
\thanks{E. Rosnes and \O. Ytrehus are with the Selmer Center, Dept. of Informatics, University of Bergen, Norway. E-mail: \{ eirik, oyvind \}@ii.uib.no.}}
\maketitle

\begin{abstract}
This paper is devoted to the finite-length analysis of turbo decoding
over the binary erasure channel (BEC). The performance of iterative
belief-propagation (BP) decoding of low-density parity-check (LDPC)
codes over the BEC can be characterized in terms of \emph{stopping
  sets}. We describe turbo decoding on the BEC which is simpler than
turbo decoding on other channels. We then adapt the concept of
stopping sets to turbo decoding and state an exact condition for
decoding failure. Apply turbo decoding until the transmitted codeword
has been recovered, or the decoder fails to progress further. Then the
set of erased positions that will remain when the decoder stops is
equal to the unique maximum-size \emph{turbo stopping set} which is
also a subset of the set of erased positions. Furthermore, we present
some improvements of the basic turbo decoding algorithm on the BEC.
The proposed improved turbo decoding algorithm has substantially
better error performance as illustrated by the given simulation
results.  Finally, we give an expression for the turbo stopping set
size enumerating function under the uniform interleaver assumption,
and an efficient enumeration algorithm of small-size turbo stopping
sets for a particular interleaver. The solution is based on the
algorithm proposed by Garello \emph{et al.}  in 2001 to compute an
exhaustive list of \emph{all} low-weight codewords in a turbo code.
\end{abstract}


\section{Introduction} \label{sec:intro}

Low-density parity-check (LDPC) codes as opposed to turbo codes have
been studied extensively on the binary erasure channel (BEC). In
\cite{lub01}, an iterative decoding algorithm for LDPC codes over the
BEC was proposed, and it was shown that this scheme approaches channel
capacity arbitrarily close. Although carefully optimized irregular
LDPC codes with iterative decoding can achieve channel capacity on the
BEC as the code length tends to infinity, there is still some
performance loss compared to maximum-likelihood (ML) decoding of a
given fixed code of finite length. Recently, some progress has been
made towards efficient ML or near ML decoding of LDPC codes over the
BEC \cite{rav05,bur04,nik04}.  The Tanner graph representation of an
LDPC code is a bipartite graph with left and right nodes. The left
nodes correspond to codeword bits. The right nodes correspond to
parity-check constraints. It is known that iterative
belief-propagation (BP) decoding fails if and only if the set of
erased bit-positions contains a \emph{stopping set} \cite{di02}. A
stopping set is a subset of the bit-positions such that the
corresponding left nodes in the Tanner graph have the property that
all neighboring nodes are connected to the set at least twice.

In this work we consider turbo decoding over the BEC. Turbo codes have
gained considerable attention since their introduction by Berrou
\emph{et al.} \cite{ber93} in 1993 due to their near-capacity
performance and low decoding complexity. Here we consider the
conventional turbo code which is the parallel concatenation of two
identical recursive systematic convolutional encoders separated by a
pseudo-random interleaver. To accurately describe turbo decoding on
the BEC we introduce the concept of a \emph{turbo stopping set}, and
we identify an exact condition for decoding failure. Assume that we
transmit codewords of a turbo code over the BEC. Apply turbo decoding
until either the codeword has been recovered, or the decoder fails to
progress further. Then the set of erased positions that will remain
when the decoder stops is equal to the unique maximum-size turbo
stopping set which is also a subset of the set of erased positions. We
also consider improved turbo decoding on the BEC. The algorithm
applies turbo decoding until the transmitted codeword is recovered, or
the decoder fails to progress further. Then, an unknown (systematic)
bit-position is identified and its value is guessed, after which turbo
decoding is applied again. Thus, the algorithm is based on guessing
bit-values in erased bit-positions when turbo decoding does not
progress further and has the same \emph{structure} as the algorithms
in \cite{nik04}.

Recently, several algorithms have been introduced to
compute the first few terms of the weight distribution of both
parallel and serial turbo codes. Both exact algorithms (e.g.,
\cite{ros03_12,gar01,per96}) and approximate algorithms (e.g.,
\cite{cro04,vil04,ber02}) have been presented. In this work we also
show that basically all {\emph{trellis-based}} algorithms can be
adapted to find the first few terms of the turbo stopping set size
enumerating function. In particular, we have considered in detail how
to adapt the algorithm by Garello \emph{et al.} introduced in
\cite{gar01} and the improved algorithm in \cite{ros03_12}. Also, an
expression for the (average) turbo stopping set size enumerating
function under the uniform interleaver assumption is presented.

Using linear programming (LP) to decode binary linear codes has
recently been considered by Feldman \emph{et al.} \cite{fel05}. See
also the seminal papers \cite{fel02,fel02_1} where LP decoding of
turbo-like codes is considered. In particular, repeat-accumulate (RA)
codes are considered. A description of LP decoding of arbitrary
concatenated codes is given in \cite[Ch.\ 6]{fel03}.  The obvious
polytope for LP decoding is the convex hull of all codewords, in which
case LP decoding is equivalent to ML decoding. However, the convex
hull has a description complexity that is exponential in the codeword
length for a general binary linear code. Thus, Feldman \emph{et al.}
\cite{fel05} proposed a relaxed polytope which contains all valid
codewords as vertices, but also additional non-codeword vertices. The
vertices of the relaxed polytope are basically what the authors called
\emph{pseudo-codewords} in \cite{fel05}.  One desirable property of
the LP decoder is the ML certificate property, i.e., when the LP
decoder outputs a codeword, it is guaranteed to be the ML codeword.
Experimental results with LDPC codes show that the performance of the
relaxed LP decoder is better than with the iterative min-sum
algorithm, but slightly worse than with iterative BP decoding.

Recently, some understanding of the performance of iterative BP
decoding of finite-length LDPC codes over general memoryless channels
have been developed. Finite graph covers of the Tanner graph and the
codes defined by these covers play an essential role in the analysis
\cite{koe05}. The low complexity of iterative BP decoding is due to
the fact that the algorithm operates \emph{locally} on the Tanner
graph of the code. This property is also the main weakness of
iterative BP decoding, since the decoder cannot distinguish if it
operates on the original Tanner graph or on any of the finite covers.
Hence, codewords in the code defined by a finite cover of the Tanner
graph will influence on iterative decoding. These codewords are
basically what Vontobel and Koetter referred to as pseudo-codewords in
\cite{koe05}.  It turns out that the set of pseudo-codewords of all
finite covers of the Tanner graph is equal to the set of points where
all entries are rational numbers from the relaxed polytope of LP
decoding of LDPC codes as introduced by Feldman \emph{et al.} in
\cite{fel05}.  A similar connection between the relaxed polytope of LP
decoding of turbo codes, as described in \cite[Ch.\ 6]{fel03}, and the
pseudo-codewords of all finite covers of the turbo code factor graph
\cite{ksc01} was established in \cite{ros06}. Furthermore, Rosnes also
showed in \cite{ros06} that there is a many-to-one correspondence
between these pseudo-codewords and turbo stopping sets in the
following sense. The support set of any pseudo-codeword, i.e., the set
of non-zero coordinates, is a turbo stopping set, and for any turbo
stopping set there is a pseudo-codeword with support set equal to the
turbo stopping set. A similar connection also holds for
pseudo-codewords of finite graph covers of Tanner graphs and stopping
sets \cite{koe05,kel05}.

For LDPC codes it has been observed that stopping sets, to some
degree, also reflect the performance of iterative decoding for other
channels than the BEC \cite{koe05,tia03}. It is therefore our hope
that the notion of turbo stopping sets also can provide some useful
insight into turbo decoding on the additive white Gaussian noise
(AWGN) channel or on other memoryless channels. In \cite{ros06},
Rosnes presented some simulation results to indicate that this may be
the case.

This paper is organized as follows. In Section~\ref{sec:decoding} we
define some basic notation and describe simplified turbo decoding on
the BEC. Section~\ref{sec:stopping} introduces the concept of a turbo
stopping set. We further give some of the basic properties and show
that turbo stopping sets characterize exactly the performance of turbo
decoding on the BEC. An improved turbo decoding algorithm on the BEC
is introduced in Section~\ref{sec:improved}, and its superiority
compared to conventional turbo decoding is illustrated by simulation
examples. Finally, in Section~\ref{sec:alg}, we consider enumeration
of small-size turbo stopping sets for a particular interleaver and
under the uniform interleaver assumption. Conclusions and a discussion
of future work are given in Section~\ref{sec:conclu}.

\section{Preliminaries} \label{sec:decoding}

In this section we introduce the channel, define some basic notation,
and describe simplified turbo decoding on the BEC.

\subsection{The BEC}

The BEC model was introduced by Elias \cite{eli55} in $1955$. The
channel has recently been used for modeling information transmission
over the Internet. The BEC is a two-input, three-output discrete
memoryless channel. Each input bit is erased with probability
$\epsilon$, or received correctly with probability $1-\epsilon$.

\subsection{Some Definitions and Basic Notation}
Let $\mathcal{C}=\mathcal{C}(K,C_a,C_b,\pi)$ denote a parallel
concatenated convolutional code (PCCC), or turbo code, with
information length $K$, constituent encoders $C_a$ and $C_b$ of rate
$R=k/n$, and interleaver $\pi$.  In this work we consider dual
termination \cite{gui94}, which implies that the length of the
interleaver $I=K+ 2\nu$ where $\nu$ is the constraint length. We
assume here that $I$ is a multiple of $k$. Let $N_a$ and $N_b$ denote
the lengths of the constituent codes. The length of the turbo code is
denoted by $N$. For an unpunctured turbo code, $N_a=N_b=I/R$ and
$N=(2/R-1)I$. In general, the values of $N_a$, $N_b$, and $N$ depend
on the puncturing pattern $P$ and the termination scheme.

\begin{figure}[htb]
\par
\psfrag{I0}{$I_0$}
\psfrag{I1}{$I_1$}
\psfrag{I2}{$I_2$}
\psfrag{I3}{$I_3$}
\psfrag{I4}{$I_4$}
\psfrag{I5}{$I_5$}
\psfrag{P0a}{$P_0^a$}
\psfrag{P2a}{$P_2^a$}
\psfrag{P4a}{$P_4^a$}
\psfrag{P1b}{$P_1^b$}
\psfrag{P3b}{$P_3^b$}
\psfrag{P5b}{$P_5^b$}
\psfrag{ast}{$\ast$}
\psfrag{psia}{$\psi_a$}
\psfrag{psib}{$\psi_b$}
\psfrag{mua}{$\mu_a$}
\psfrag{mub}{$\mu_b$}
\begin{center}
\includegraphics[width=3.5in,height=2.0in]{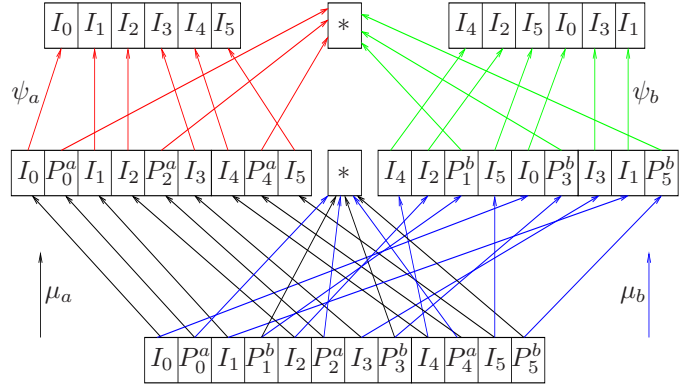}
\end{center}
\caption{\label{figur_mapping} {Graphical representation of the mappings $\mu_a$, $\mu_b$, $\psi_a$, and $\psi_b$ for the turbo code from Example~\ref{example1}.}}
\end{figure}

We will now define some useful mappings, but we advise the reader that
the formal definitions below will be easier to understand after taking
a look at Fig.~\ref{figur_mapping}.

Define two mappings $\mu_a: \mathcal{Z}_N \rightarrow
\mathcal{Z}_{N_a} \cup \{\ast\}$ and $\mu_b: \mathcal{Z}_N
\rightarrow{Z}_{N_b} \cup \{\ast\}$ where
$\mathcal{Z}_N=\{0,1,\dots,N-1\}$ for a positive integer $N$. The
mapping $\mu_a$ gives the index in the first constituent codeword of
the turbo codeword index if such a relation exists, or $\ast$ if not.
Similarly, the mapping $\mu_b$ gives the index in the second
constituent codeword of the turbo codeword index if such a relation
exists, or $\ast$ if not. Note that for turbo codeword indices that
correspond to systematic bits, the interleaver is used to get the
correct constituent codeword index for the second encoder.

Define two mappings $\psi_a: \mathcal{Z}_{N_a} \rightarrow
\mathcal{Z}_{I} \cup \{\ast\}$ and $\psi_b: \mathcal{Z}_{N_b}
\rightarrow{Z}_{I} \cup \{\ast\}$. The mapping $\psi_a$ gives the
systematic sequence index of the first constituent codeword index if
such a relation exists, or $\ast$ if not. Similarly, the mapping
$\psi_b$ gives the interleaved systematic sequence index of the second
constituent codeword index if such a relation exists, or $\ast$ if
not.

\begin{example} \label{example1}
Consider a turbo code composed of two identical nominal rate-$1/2$
constituent convolutional codes. The interleaver length $I=6$ and
parity bits from the two constituent encoders are punctured
alternatively to create a nominal rate-$1/2$ turbo code. The lengths
of the first and second constituent codes are $N_a=N_b=9$. The
interleaver $\pi$ is defined by $\{3,5,1,4,0,2\}$ (i.e., $0
\rightarrow 3$, $1 \rightarrow 5$, and so on).  The ordering of bits
in the turbo codeword is
\begin{displaymath}
I_0 P_0^a I_1 P_1^b I_2 P_2^a I_3 P_3^b I_4 P_4^a I_5 P_5^b
\end{displaymath}
where $I_i$ and $P_i^a$ denote the $i$th systematic and parity bit
from the first constituent code, respectively, and where $P_i^b$
denotes the $i$th parity bit from the second constituent code. The
mappings $\mu_a$, $\mu_b$, $\psi_a$, and $\psi_b$ are depicted
graphically in Fig.~\ref{figur_mapping}.
\end{example}

\subsection{Turbo Decoding on the BEC} \label{sec:turbodec}

The aim of turbo decoding on the BEC is to find a set of paths through
each constituent code trellis that is consistent with the received
sequence. The decoding starts with a set of all paths and iteratively
eliminates those that are inconsistent. This iterative process
continues until either there is only one possible path left in each
constituent trellis (successful decoding), or there is no change from
one iteration step to the next. We will describe the basic algorithm,
but remark that the BEC version of turbo decoding allows more
efficient implementations. For simplicity, we omit the details.

Let $\mathcal{T}^x_{\rm info}$ denote an \emph{information
  bit-oriented trellis} \cite{ros02_2,mce96,cha96} for constituent
code $C_x$, $x=a,b$. The trellis $\mathcal{T}^x_{\rm info}$ is
time-variant, but periodic, and has two edges out of each vertex. For
details, see Section~\ref{sec:conv}. The number of trellis depths in
$\mathcal{T}^x_{\rm info}$ is $I+1$, i.e., there is one trellis
section for each information bit.  Below, $\bar{x}$ denotes the
complement of $x$, $x=a,b$.

Like turbo decoding for an AWGN channel, the constituent decoders work
with a forward and a backward pass through the constituent code
trellis $\mathcal{T}^x_{\rm info}$, and during these passes the state
metric $\alpha_x^{(i)}(v,j)$ (resp.\ $\beta_x^{(i)}(v,j)$) for vertex
$v$ at trellis depth $j$, at the $i$th iteration, is updated.  For the
BEC, however, the state metrics are boolean and define forward (resp.\
backward) paths that are consistent with the current estimate of the
transmitted sequence $\hat{\bf c}=(\hat{c}_0,\dots,\hat{c}_{N-1})$.
The decoding for constituent code $C_x$ is performed as follows.

Initially, we set $\alpha_x^{(1)}(0,0) = \beta_x^{(1)}(0,I) =
\beta_x^{(0)}(0,I)={\bf true}$; $\beta_x^{(0)}(v,j) = {\bf true}$ for
all vertices $v$ at trellis depth $j$ in $\mathcal{T}^x_{\rm info}$,
$j=1,\dots,I-1$; and $\alpha_x^{(1)}(v,0) = \beta_x^{(1)}(v,I) =
\beta_x^{(0)}(v,I)={\bf false}$ for every non-zero vertex $v$ at
trellis depths $0$ and $I$, respectively, in $\mathcal{T}^x_{\rm
  info}$. Finally, set the estimate of the transmitted sequence
$\hat{\bf c}$ equal to the received sequence.

Then the forward pass calculates, for $j=1,\dots,I$ and for each
vertex $v'$ at trellis depth $j$ in $\mathcal{T}^x_{\rm info}$,
$\zeta_j(v')$ and subsequently $\alpha_x^{(i)}(v',j) =
(\beta_x^{(i-1)}(v',j)\; {\rm AND}\; \zeta_j(v'))$. Here $\zeta_j(v')$
is a boolean variable which is true if there exists, in the $(j-1)$th
trellis section of $\mathcal{T}^x_{\rm info}$, an edge consistent with
$\hat{\bf c}$, with right vertex $v'$, and left vertex $v$ such that
$\alpha_x^{(i)}(v,j-1)= {\bf true}$. Initially, prior to the forward
pass, we set $\alpha_x^{(i)}(v,0)=\alpha_x^{(1)}(v,0)$ for all
vertices $v$ at trellis depth $0$ in $\mathcal{T}^x_{\rm info}$.

In a similar way the backward pass calculates $\beta_x^{(i)}(v,j)$ for
all vertices $v$ at trellis depth $j$ in $\mathcal{T}^x_{\rm info}$,
$j=0,\dots,I-1$. Finally, the estimate $\hat{\bf c}$ of the
transmitted sequence is updated so that only information values
consistent with legal edges remain, and control is passed to the other
constituent decoder.

The codeword is said to be \emph{recovered} if $\hat{c}_j \neq \star$,
where $\star$ denotes an erasure, for all $j$ such that $\mu_a(j) \neq
\ast$ and $\psi_a(\mu_a(j)) \neq \ast$, $j=0,\dots,N-1$. The decoder
is said to \emph{fail to progress further} if, for $x=a,b$, the state
metrics $\alpha_x^{(l)}(v,j)$ and $\beta_x^{(l)}(v,j)$, for some
positive integer $l>1$, are equal to the state metrics
$\alpha_x^{(l-1)}(v,j)$ and $\beta_x^{(l-1)}(v,j)$, respectively, for
all vertices $v$ at trellis depth $j$ in $\mathcal{T}^x_{\rm info}$,
$j=1,\dots,I-1$.

\section{Turbo Stopping Sets} \label{sec:stopping}
In this section we will introduce the concept of a turbo stopping set.
A turbo stopping set is the equivalent of an LDPC stopping set when
turbo decoding and \emph{not} iterative BP decoding is performed.

\begin{definition} \label{def_1}
Let $\mathcal{C}$ denote a given PCCC with interleaver $\pi$.  A set
$\mathcal{S} = \mathcal{S}(\pi) \subseteq \{0,\dots,N-1\}$ is a turbo
stopping set if and only if there exist two linear subcodes $\bar{C}_a
\subseteq C_a \subseteq \{0,1 \}^{N_a}$ and $\bar{C}_b \subseteq C_b
\subseteq \{0,1 \}^{N_b}$ of dimension $>0$ with support sets
$\chi(\bar{C}_a)$ and $\chi(\bar{C}_b)$, respectively, such that
\begin{align} \label{eq:stoppingset}
\chi(\bar{C}_a) &= \mu_a(\mathcal{S}) \setminus \{\ast\} \notag \\
\chi(\bar{C}_b) &= \mu_b(\mathcal{S}) \setminus \{\ast\} \notag \\
\pi(\psi_a(\chi(\bar{C}_a)) \setminus \{\ast\}) &= \psi_b(\chi(\bar{C}_b)) \setminus \{\ast\}.
\end{align}
The \emph{size} of a turbo stopping set $\mathcal{S}$ is its cardinality.
\end{definition}

The lemmas below state some of the properties of a turbo stopping set.

\begin{lemma} \label{lemma1}
Let $\mathcal{C}$ denote a given PCCC with interleaver $\pi$. The
support set of any non-zero codeword from $\mathcal{C}$ is a turbo
stopping set of size equal to the Hamming weight of the given
codeword. Thus, the minimum turbo stopping set size is upper-bounded
by the minimum Hamming weight.
\end{lemma}

\begin{proof}
Denote the turbo codeword by $\mathbf{c}$ and the corresponding first
and second constituent codewords by $\mathbf{c}_a$ and $\mathbf{c}_b$,
respectively. Furthermore, let $\bar{C}_a =
\{\mathbf{c}_a,\mathbf{0}_{N_a} \}$ and $\bar{C}_b =
\{\mathbf{c}_b,\mathbf{0}_{N_b} \}$ where $\mathbf{0}_{N_x}$ denotes
an all-zero sequence of length $N_x$, $x=a,b$. The result follows
immediately from Definition~\ref{def_1}, since
$\chi(\bar{C}_a)=\mu_a(\chi(\mathbf{c})) \setminus \{\ast\}$,
$\chi(\bar{C}_b)=\mu_b(\chi(\mathbf{c})) \setminus \{\ast\}$, and
$\pi(\psi_a(\mu_a(\chi(\mathbf{c})) \setminus \{\ast\}) \setminus
\{\ast\}) = \psi_b(\mu_b(\chi(\mathbf{c})) \setminus \{\ast\})
\setminus \{\ast\}$.
\end{proof}

\begin{lemma} \label{lemma2}
Let $\mathcal{C}$ denote a given PCCC with interleaver $\pi$, and let
$\mathcal{S}=\mathcal{S}(\pi)$ denote a turbo stopping set. If
$\bar{C}_a$ and $\bar{C}_b$ can both be decomposed into \emph{direct
  sums} of linear subcodes of dimension $1$ with disjoint support
sets, then $\mathcal{S}$ is the support set of a turbo codeword of
Hamming weight $|\mathcal{S}|$. The converse is also true.
\end{lemma}

\begin{proof}
Assume that $\bar{C}_a$ and $\bar{C}_b$ can both be decomposed into
direct sums of linear subcodes of dimension $1$ with disjoint support
sets. In more detail,
\begin{displaymath}
\bar{C}_a = \{\mathbf{c}_a^{(0)},\mathbf{0}_{N_a} \} + \cdots + \{\mathbf{c}_a^{(p)},\mathbf{0}_{N_a} \} \text{ and }
\bar{C}_b = \{\mathbf{c}_b^{(0)},\mathbf{0}_{N_b} \} + \cdots + \{\mathbf{c}_b^{(q)},\mathbf{0}_{N_b} \}
\end{displaymath}
where $p$ and $q$ are non-negative integers. The codewords
$\mathbf{c}_a = \sum_{i=0}^p \mathbf{c}_a^{(i)}$ and $\mathbf{c}_b =
\sum_{i=0}^q \mathbf{c}_b^{(i)}$ have support sets
$\chi(\mathbf{c}_a)=\chi(\bar{C}_a)$ and
$\chi(\mathbf{c}_b)=\chi(\bar{C}_b)$, respectively, since the support
sets of the direct sum subcodes are disjoint. Since the two sets
$\pi(\psi_a(\mu_a(\mathcal{S}) \setminus \{\ast\}) \setminus
\{\ast\})=\pi(\psi_a(\chi(\mathbf{c}_a)) \setminus \{\ast\})$ and
$\psi_b(\mu_b(\mathcal{S}) \setminus \{\ast\}) \setminus
\{\ast\}=\psi_b(\chi(\mathbf{c}_b)) \setminus \{\ast\}$ are equal
(from Definition~\ref{def_1}), there exists a turbo codeword
$\mathbf{c}$ with first and second constituent codewords
$\mathbf{c}_a$ and $\mathbf{c}_b$, respectively. Thus, the turbo
stopping set is the support set of a turbo codeword of Hamming weight
$|\mathcal{S}|$, since $\chi(\mathbf{c})=\mathcal{S}$.

We prove the converse using the following argument. The turbo stopping
set $\mathcal{S}$ is the support set of a turbo codeword $\mathbf{c}$.
The corresponding first and second constituent codewords are denoted
by $\mathbf{c}_a$ and $\mathbf{c}_b$, respectively. Then, there exist
subcodes $\bar{C}_a=\{\mathbf{c}_a,\mathbf{0}_{N_a} \} \subset C_a$
and $\bar{C}_b=\{\mathbf{c}_b,\mathbf{0}_{N_b} \} \subset C_b$ both of
dimension $1$ which satisfy the constraints in (\ref{eq:stoppingset})
with $\mathcal{S}=\chi(\mathbf{c})$, and the result follows. (The
codewords $\mathbf{c}_a$ and $\mathbf{c}_b$ may or may not be further
decomposed.)
\end{proof}

\begin{construction} \label{construct:1}
Let $T_a=(V_a,E_a)$ and $T_b=(V_b,E_b)$ denote two \emph{arbitrarily}
chosen Tanner graphs for the first and second constituent codes,
respectively. The node set $V_x$ can be partitioned into three
disjoint subsets $V_x^s$, $V_x^p$, and $V_x^c$, corresponding to
systematic bits, parity bits, and parity-check equations,
respectively, where $x=a,b$. There is an edge $e \in E_x$ connecting a
node $v_x^v \in V_x^v = V_x^s \cup V_x^p$ to a node $v_x^c \in V_x^c$
if and only if the first ($x=a$) or second ($x=b$) constituent
codeword bit represented by $v_x^v$ is checked by the parity-check
equation represented by $v_x^c$. Next, define
\begin{displaymath}
E'_b = \{(v_{a,j}^s,v_b^c):v_b^c \in V_b^c,(v_{b,\pi(j)}^s,v_b^c) \in E_b, j=0,\dots,I-1 \} \cup \{(v_b^p,v_b^c): v_b^p \in V_b^p, v_b^c \in V_b^c, (v_b^p,v_b^c) \in E_b \}.
\end{displaymath}
Then, the graph $T=(V_a \cup V_b^p \cup V_b^c,E_a \cup E'_b)$ is a
Tanner graph for the turbo code $\mathcal{C}$.
\end{construction}

We remark that for a given binary linear code there exist in general
several full-rank parity-check matrices, and thus several (distinct)
Tanner graphs. The minimum stopping set size is in general a function
of the Tanner graph \cite{sch05_01,web05,hol05}. Note that
Construction~\ref{construct:1} gives a specific class of Tanner graphs
for a turbo code that is a proper subset of the class of all Tanner
graphs of the given turbo code. We will consider this specific class
of Tanner graphs below.

\begin{lemma}
Let $\mathcal{C}$ denote a given PCCC with interleaver $\pi$. For any
turbo stopping set $\mathcal{S}=\mathcal{S}(\pi)$ there is an LDPC
stopping set of cardinality $|\mathcal{S}|$ in any Tanner graph $T$
for $\mathcal{C}$ within the class of Tanner graphs from
Construction~\ref{construct:1}.
\end{lemma}

\begin{proof}
We use the notation introduced in Construction~\ref{construct:1}. The
set $\{v^v_{x,j}: j \in \mu_x(\mathcal{S}) \setminus \{\ast\}\}$,
where $v^v_{x,j}$ denotes the $j$th node in $V_x^v$, is an LDPC
stopping set in any Tanner graph $T_x$ of $C_x$, since
$\mu_x(\mathcal{S}) \setminus \{\ast\}$ is the support set
$\chi(\bar{C}_x)$ (from Definition~\ref{def_1}) of some subcode
$\bar{C}_x$ of $C_x$ of dimension $>0$. We have here used the fact
that the variables nodes corresponding to the support set of a
non-zero codeword constitute an LDPC stopping set.  Furthermore, the
set
\begin{equation} \label{eq:set}
\left(\{v_{a,j}^v: j \in \mu_a(\mathcal{S}) \setminus \{\ast\} \} \cup \{v_{b,j}^v: j \in \mu_b(\mathcal{S}) \setminus \{\ast\} \} \right) \setminus V^s_b
\end{equation}
is an LDPC stopping set in the Tanner graph $T$ of the turbo code
$\mathcal{C}$ due to the last condition in Definition~\ref{def_1}. The
cardinality of the LDPC stopping set in (\ref{eq:set}) is
$|\mathcal{S}|$.
\end{proof}

Note that the converse is not necessarily true (i.e., for an LDPC
stopping set in any Tanner graph $T$ for $\mathcal{C}$ within the
class of Tanner graphs from Construction~\ref{construct:1}, there is
not necessarily a turbo stopping set). Thus, iterative BP decoding
using a Tanner graph within the class of Tanner graphs of a turbo code
from Construction~\ref{construct:1} is inferior to turbo decoding on
the BEC. The following theorem states an exact condition for decoding
failure.

\begin{theorem}
Let $\mathcal{C}$ denote a given PCCC with interleaver $\pi$ that we
use to transmit information over the BEC. The received vectors are
decoded using turbo decoding until either the codeword has been
recovered, or the decoder fails to progress further. Then the set of
erased positions that will remain when the decoder stops is equal to
the unique maximum-size turbo stopping set which is also a subset of
$\mathcal{E}$, where $\mathcal{E}$ denotes the subset of erased
positions.
\end{theorem}

\begin{proof}
The proof given here is inspired by the proof given by Di \emph{et
  al.} in \cite[Lemma~1.1]{di02} in the context of an LDPC stopping
set.  Let $\mathcal{S}$ be a turbo stopping set contained in
$\mathcal{E}$.  The claim is that the basic turbo decoder cannot
determine the bits corresponding to the positions in the turbo
stopping set $\mathcal{S}$.  Assume that all other bits are known.
Turbo decoding starts by activating the first constituent decoder.
For the first constituent decoder, the forward-backward algorithm will
determine $|\bar{C}_a|>1$ possible paths through the trellis.  The
support set of these possible paths is equal to $\mu_a(\mathcal{S})
\setminus \{\ast\}$, and they are all equally likely. Consequently, no
additional codeword bits can be determined.  Thus, the extrinsic
probability distributions for systematic bits in positions in
$\psi_a(\mu_a(\mathcal{S}) \setminus \{\ast\}) \setminus \{\ast\}$ are
uniform. For the second constituent decoder, the forward-backward
algorithm will determine $|\bar{C}_b|>1$ possible paths through the
trellis. The support set of these possible paths is equal to
$\mu_b(\mathcal{S}) \setminus \{\ast\}$ and, since the \emph{a priori}
probability distributions for systematic bits in positions in
$\pi(\psi_a(\mu_a(\mathcal{S}) \setminus \{\ast\}) \setminus
\{\ast\})$ are uniform, and the two sets
$\pi(\psi_a(\mu_a(\mathcal{S}) \setminus \{\ast\}) \setminus
\{\ast\})$ and $\psi_b(\mu_b(\mathcal{S}) \setminus \{\ast\})
\setminus \{\ast\}$ are equal (see Definition~\ref{def_1}), all paths
are equally likely.  Consequently, no additional codeword bits can be
determined. Thus, the extrinsic probability distributions for
systematic bits in positions in $\psi_b(\mu_b(\mathcal{S}) \setminus
\{\ast\}) \setminus \{\ast\}$ are uniform. One iteration of the basic
turbo decoding algorithm has been performed and no additional codeword
bits have been determined. Since there is no new information available
to the first constituent decoder, no additional bits will be
determined in the next round of turbo decoding either. It follows that
the decoder cannot determine the bits corresponding to the positions
in the unique maximum-size turbo stopping set which is also a subset
of $\mathcal{E}$. Note that there is a unique maximum-size turbo
stopping set which is also a subset of $\mathcal{E}$, since the union
of two turbo stopping sets is also a turbo stopping set. Conversely,
if the decoder terminates at a set $\mathcal{S}$, then there will
exist subcodes $\bar{C}_a \subseteq C_a$ and $\bar{C}_b \subseteq C_b$
of dimension $>0$ with support sets $\mu_a(\mathcal{S}) \setminus
\{\ast\}$ and $\mu_b(\mathcal{S}) \setminus \{\ast\}$, respectively.
Since the turbo decoder terminates, the two sets
$\pi(\psi_a(\mu_a(\mathcal{S}) \setminus \{\ast\}) \setminus
\{\ast\})$ and $\psi_b(\mu_b(\mathcal{S}) \setminus \{\ast\})
\setminus \{\ast\}$ are equal. From Definition~\ref{def_1}, it follows
that $\mathcal{S}$ is a turbo stopping set and, since no erased
bit-positions contained in a turbo stopping set can be determined by
the turbo decoder, it must be the maximum-size turbo stopping set
which is also a subset of $\mathcal{E}$.
\end{proof}

\subsection{A $(155,64,18)$ Turbo Code} \label{sec:155turbo}

In \cite{tan01}, a particularly nice $(3,5)$-regular LDPC code of
length $155$, dimension $64$, and minimum Hamming distance $20$ was
constructed. The underlying Tanner graph has girth $8$ which makes the
code an excellent candidate for iterative decoding. This is the reason
behind the selected code parameters.

\begin{figure*}[htb]
\par
\begin{center}
\includegraphics[width=6.5in,height=2.0in]{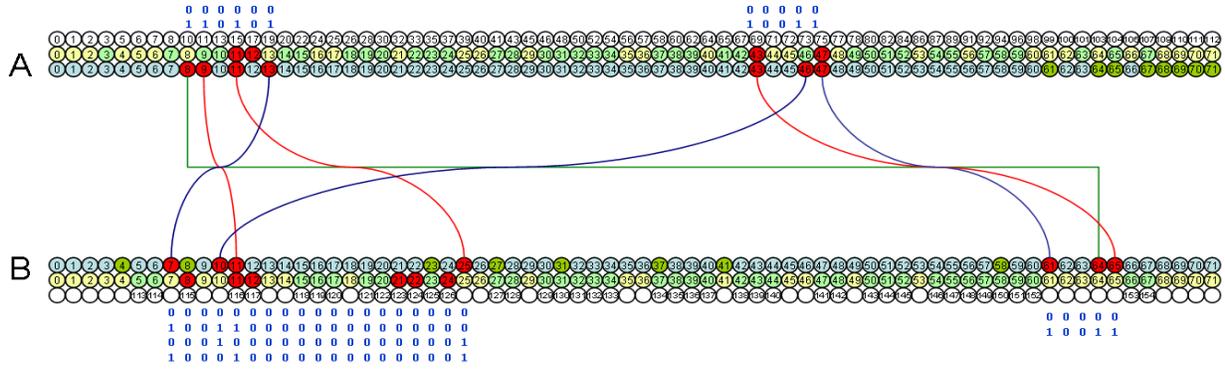}
\end{center}
\caption{\label{figur1} {Turbo stopping set of size $17$ for the example $(155,64,18)$ turbo code. The direct sum of the two upper (lower) sets of binary vectors (extended to length $I$ with zeros) is the \emph{systematic part} of the subcode $\bar{C}_a$ ($\bar{C}_b$). }}
\end{figure*}

The turbo code is obtained by puncturing of a nominal rate-$1/3$ turbo
code with nominal rate-$1/2$, constraint length $\nu=4$ constituent
codes defined by the parity-check matrix $\mathbf{H}(D) =
(1+D+D^2+D^4\;1+D^3+D^4)$.  The last polynomial which is irreducible
and primitive has been chosen as the parity polynomial making the
constituent encoders recursive.  The information block size is $64$,
the interleaver length is $72$ due to dual termination \cite{gui94},
and the interleaver is a dithered relative prime (DRP) interleaver
\cite{cro04,cro03}. The puncturing pattern is designed using the
algorithm in \cite{ros05isit}.  The minimum distance of $18$ of the
code has been computed using the algorithm in \cite{ros03_12}. For
this code there exists a turbo stopping set of size $17$. The turbo
stopping set is depicted in Fig.~\ref{figur1}. In Fig.~\ref{figur1},
the three upper rows of nodes correspond to the first constituent
encoder, while the three remaining rows of nodes correspond to the
second constituent encoder. The nodes in row number $5(i-1)+1$ give
the bit-position in the turbo codeword of the corresponding systematic
($i=1$) and parity ($i=2$) bits. The nodes in row number $3(i-1)+2$
and $i+2$ correspond to parity and systematic bits, respectively, from
constituent encoder $i$, $i=1,2$. The blue and dark green nodes
correspond to systematic bits. The blue nodes are information bits
(can be assigned freely), while the dark green nodes are redundant
systematic bits (can \emph{not} be assigned freely). The red and
yellow nodes correspond to erased bits and punctured (parity) bits,
respectively. The light green nodes correspond to parity bits. The
arrows in between the upper three rows and the remaining three rows
correspond to interleaving of erased information bits. The remaining
part of the interleaver is of no concern for the following discussion.
In fact, possible choices for $\bar{C}_a$ and $\bar{C}_b$ are the
linear subcodes spanned by
\begin{align}
\mathbf{c}_a^{(1)} = (&0\diamond,\dots,00,1\diamond,10,00,11,01,1\diamond,00,\dots,00,00,0\diamond, 0\diamond,00,00,0\diamond,\dots,0\diamond) \notag \\
\mathbf{c}_a^{(2)} =(&0\diamond,\dots,00,0\diamond,00,00,00,00,0\diamond,00,\dots,00,11,0\diamond,0\diamond,10,11,0\diamond,\dots,0\diamond)
\end{align}
and
\begin{align}
\mathbf{c}_b^{(1)} =(&0\diamond,\dots,00,1\diamond,01,0\diamond,0\diamond,10,01,0\diamond,0\diamond,00,00,00,0\diamond,00,00,01,01,00,01,1\diamond,0\diamond,\dots,00,0\diamond,0\diamond,0\diamond, \notag \\
& 0\diamond,0\diamond,00,\dots,0\diamond) \notag \\
\mathbf{c}_b^{(2)} =(&0\diamond,\dots,00,0\diamond,00,0\diamond,1\diamond,01,01,0\diamond,0\diamond,00,00,00,0\diamond,00,00,01,01,00,01,1\diamond,0\diamond,\dots,00,0\diamond,0\diamond,0\diamond, \notag \\
& 0\diamond,0\diamond,00,\dots,0\diamond) \notag \\
 \mathbf{c}_b^{(3)} =(&0\diamond,\dots,00,0\diamond,00,0\diamond,0\diamond,00,00,0\diamond,0\diamond,00,00,00,0\diamond,00,00,00,00,00,00,0\diamond,0\diamond,\dots,00,1\diamond,0\diamond,0\diamond,\notag \\
& 1\diamond,1\diamond,00,\dots,0\diamond),
\end{align}
respectively.  The symbol $\diamond$ indicates that the bit-position
has been punctured. From Fig.~\ref{figur1} we get
\begin{align}
&\mathcal{S} =\{10,11,15,16,18,19,69,70,73,75,76,115,116,117,123,124,126\} \notag \\
&\mu_a(\mathcal{S}) = \{10,11,15,16,18,19,69,70,73,75,76,\ast\} \notag \\
&\mu_b(\mathcal{S}) = \{9,11,13,14,15,17,32,34,38,39,101,104,105,\ast\} \notag \\
&\psi_a(\mu_a(\mathcal{S}) \setminus \{\ast\}) = \{8,9,11,13,43,46,47,\ast\} \notag \\
&\psi_b(\mu_b(\mathcal{S}) \setminus \{\ast \}) = \{7,10,11,25,61,64,65,\ast\}.
\end{align}
Furthermore, it holds that $\chi(\bar{C}_a)=\mu_a(\mathcal{S})
\setminus \{\ast\}$, $\chi(\bar{C}_b)=\mu_b(\mathcal{S}) \setminus
\{\ast\}$, and $\pi(\psi_a(\chi(\bar{C}_a)) \setminus \{\ast\}) =
\psi_b(\chi(\bar{C}_b)) \setminus \{\ast\}$, which shows that
$\mathcal{S}$ is a turbo stopping set.

The turbo stopping set depicted in Fig.~\ref{figur1} is given as an
example to illustrate the concept of a turbo stopping set. Actually,
it is possible to find an exhaustive list of all turbo stopping sets
of size less than some threshold using a modification of the algorithm
in \cite{ros03_12}. The details of the algorithm are outlined in
Section~\ref{sec:alg}. For this code there are $2$ turbo stopping sets
of size $16$ and $13$ of size $17$.

\subsection{A $(201,64,21)$ Turbo Code}
From Fig.~\ref{figur1} it is apparent that some of the punctured bits
can be reinserted without increasing the size of the depicted turbo
stopping set. In fact, $27$ out of the $31$ punctured parity bits from
the first constituent code can be reinserted. For the second
constituent code, $19$ out of the $30$ punctured parity bits can be
reinserted. The minimum distance of $21$ of the resulting turbo code
has been determined by the algorithm in \cite{ros03_12}. Note that
this is a \emph{constructed} example, since it is possible to design a
puncturing pattern of weight $15$ that will give a length $201$ and
dimension $64$ turbo code of minimum distance $25$ using the same
mother code and the algorithm in \cite{ros05isit}. The minimum
distance of the unpunctured mother code is $27$.

\subsection{Remarks}
The code considered in Section~\ref{sec:155turbo} is \emph{not} a rare
example in the sense of having turbo stopping sets of size smaller
than the minimum distance. We have found several examples of excellent
turbo codes with this property. For further examples see
Section~\ref{sec:numerical1}.

\section{Improved Turbo Decoding} \label{sec:improved}

In \cite{nik04}, Pishro-Nik and Fekri consider improved iterative BP
decoding of LDPC codes on the BEC. When standard iterative BP decoding
fails, the improved algorithm chooses one of the unknown variables
nodes and guesses its value. The decoding continues until either the
transmitted codeword is recovered, the decoder does not progress
further, or the decoder reaches an \emph{inconsistency}. The decoder
is said to reach an inconsistency if all the variables nodes connected
to a check node are known, but the check node is not satisfied. If the
decoder has reached an inconsistency, then the value of the guessed
variable node is changed and the decoding is repeated. This time the
decoder will not reach an inconsistency, but the decoding can stop
again. More sophisticated algorithms for improved iterative BP
decoding of LDPC codes on the BEC were recently proposed by Ravisankar
and Fekri in \cite{rav05}. These algorithms improve upon the
algorithms in \cite{nik04}. Furthermore, the algorithms in
\cite{rav05} use the Tanner graph representation of the LDPC code
actively to identify \emph{equivalent} bit-positions when iterative BP
decoding has stopped in a stopping set. In this context, two
bit-positions are equivalent if and only if the knowledge of one of
the bits implies the knowledge of the other bit after a series of
message-passing in the subgraph of the Tanner graph composed of all
unknown variable nodes and the neighboring check nodes of degree $2$.

Improved turbo decoding, the pseudo-code of which is given below, has
the same \emph{structure} as the algorithm in \cite{nik04}. In more
detail, it is based on guessing unknown systematic bit-positions and
then decode until either the transmitted codeword is recovered, the
decoder does not progress further, or the decoder reaches some kind of
inconsistency. In the context of turbo decoding, however, the unknown
bit-position to guess can be chosen more efficiently than in the
context of improved iterative BP decoding of LDPC codes, since the
forward and backward passes on the constituent trellises give
information on legal paths. Bit-position selection is considered in
more detail in Section~\ref{sec:turbo_improved}.

In the pseudo-code below $\gamma_{x,j}^{(l)}$, $x=a,b$, denotes the
number of vertices of $\mathcal{T}^x_{\rm info}$ at trellis depth $j$
that are legal after the $l$th iteration.

\noindent\textbf{step 1 (Initialization)}:
\begin{enumerate}
\item Choose the maximal number of iterations $l_{\rm max}$.
\item Initialize $J$ and $T$ with zero.
\end{enumerate}

\noindent\textbf{step 2 (Original turbo decoding; bit-position selection)}:
\begin{enumerate}
\item Run turbo decoding as described in Section~\ref{sec:turbodec}, with a maximum of $l_{\rm max}$ iterations, until either the codeword is recovered, or the decoder fails to progress further. Let $l_J \leq l_{\rm max}$ denote the  number of iterations carried out and increment $T$ with $l_J$. 
\item If $T=l_{\rm max}$, or the codeword is recovered, make a decision based on $\hat{\mathbf{c}}$, and terminate the algorithm.
\item Compute $\gamma_{x,0}^{(T)},\dots,\gamma_{x,I}^{(T)}$ for $x=a,b$ and select a bit-position $v_T$ to guess based on these values. See Section~\ref{sec:turbo_improved} below for details.
\item Initialize a list $\mathcal{L}$ with the \emph{ordered} sets $\{(v_T,0)\}$ and $\{(v_T,1)\}$.
\item Initialize $\mathcal{M}_0$ and $\mathcal{M}_1$ with
  $\hat{\mathbf{c}}$.
\end{enumerate}

\noindent\textbf{step 3 (Performing additional iterations; bit-position selection)}:
\begin{enumerate}
\item Increment $J$.
\item $L=\{(L_1^{(0)},L_2^{(0)}),\dots,(L_1^{(|L|-1)},L_2^{(|L|-1)})\}$  is chosen and removed from  $\mathcal{L}$.
\item  Initialize $\hat{\mathbf{c}}$ with $\mathcal{M}_{L_2^{(0)},\dots,L_2^{(|L|-1)}}$ and set $\hat{c}_{l}$ equal to $L_2^{(j)}$ where $\psi_a(\mu_a(l))=L_1^{(j)}$ and $j=|L|-1$.
\item Run turbo decoding  as described in Section~\ref{sec:turbodec} with a maximum of $l_{\rm max}-T$ iterations, but without assigning values to $\hat{\mathbf{c}}$ in the initialization step of the algorithm, 
  until the decoder fails to progress further, or there is a trellis depth $j$ in which \emph{all} vertices in $\mathcal{T}^x_{\rm info}$ have false forward or backward state metrics for some $x$, $x=a,b$. In this  case the decoder is said to have reached an \emph{inconsistency}.  Let $l_J \leq l_{\rm max}-T$ denote the number of iterations carried out and increment $T$ with $l_J$. 
\item If  $T=l_{\rm max}$, or the decoder has not reached an inconsistency and $\hat{c}_j \neq \star$ for all $j$ such that $\mu_a(j) \neq \ast$ and $\psi_a(\mu_a(j)) \neq \ast$, $j=0,\dots,N-1$, make a decision based on $\hat{\mathbf{c}}$, and terminate the algorithm.
\item Delete $\mathcal{M}_{L_2^{(0)},\dots,L_2^{(|L|-1)}}$.
\item If  the decoder has not reached an inconsistency, perform the following.
\begin{enumerate}
\item Compute $\gamma_{x,0}^{(T)},\dots,\gamma_{x,I}^{(T)}$ for $x=a,b$ and select a bit-position $v_T$ to guess based on these values. See Section~\ref{sec:turbo_improved} below for details.
\item Add the two elements $L_0= L \cup \{(v_T,0)\}$ and $L_1= L \cup \{(v_T,1)\}$ to the list $\mathcal{L}$, and initialize $\mathcal{M}_{L_{0,2}^{(0)},\dots,L_{0,2}^{(|L_0|-1)}}$ and $\mathcal{M}_{L_{1,2}^{(0)},\dots,L_{1,2}^{(|L_1|-1)}}$ with $\hat{\mathbf{c}}$.
\end{enumerate}
\end{enumerate}

\noindent\textbf{step 4 (Repeating)}: Repeat step 3.
\vspace{1.0ex}

The list $\mathcal{L}$ in improved turbo decoding can be implemented
as a \emph{last-in first-out} queue or as a \emph{first-in first-out}
queue. The first-in first-out implementation requires more memory than
the last-in first-out implementation. The efficiency of an improved
decoding algorithm with the above structure is very dependent on the
selection of bit-positions to guess.

\subsection{Bit-Position Selection for Improved Turbo Decoding} \label{sec:turbo_improved}

When turbo decoding fails to progress further, the unknown
bit-positions constitute a turbo stopping set. Thus, guessing a
bit-position in $\chi(\bar{C}_x)$ will \emph{free} at least one
additional bit. Some of the positions in $\chi(\bar{C}_x)$ can be
determined from the numbers $\gamma_{x,j}^{(l)}$ of legal vertices at
trellis depth $j$ in $\mathcal{T}^x_{\rm info}$ ($l$ is the iteration
number). In particular, the $j$th systematic bit is unknown if
$\gamma_{a,j}^{(l)}=1$ and $\gamma_{a,j+1}^{(l)}=2$, or
$\gamma_{a,j+1}^{(l)}=1$ and $\gamma_{a,j}^{(l)}=2$. When selecting a
bit-position to guess we would also like to free as many unknown
positions as possible. We propose the following bit-position selection
algorithm.
\begin{enumerate}
\item For $x=a,b$ do the following.
\begin{enumerate}
\item Let $l_x$ be the number of indices $j$ with the property that $\gamma_{x,j}^{(T)}=1$ and $\gamma_{x,j+1}^{(T)}=2$.
\item Let $w_{x,f}$ be the largest non-negative integer such that there exists an index $f_x$ with the property that $\gamma_{x,f_x-1}^{(T)}=1$ and $\gamma_{x,f_x}^{(T)}=\cdots=\gamma_{x,f_x+w_{x,f}}^{(T)}=2$. 
\item Let $w_{x,r}$ be the largest non-negative integer such that there exists an index $r_x$ with the property that $\gamma_{x,r_x+1}^{(T)}=1$ and $\gamma_{x,r_x}^{(T)}=\cdots=\gamma_{x,r_x-w_{x,r}}^{(T)}=2$. 
\item Let $w_x = \max(w_{x,f},w_{x,r})$.
\end{enumerate}
\item If $l_a > l_b$, or $l_a=l_b$ and $w_a \geq w_b$, then set $x=a$. Otherwise, set $x=b$.
\item If $w_{x,f} \geq w_{x,r}$, then $v_T$ is set equal to $f_x-1$ if $x=a$ and $\pi^{-1}(f_x-1)$ if $x=b$. Otherwise, $v_T$ is set equal to $r_x$ if $x=a$ and $\pi^{-1}(r_x)$ if $x=b$.
\end{enumerate}

\subsection{Remarks}

We remark that several variations of the above bit-position selection
algorithm are possible. In particular, one could use both the number
of legal edges in each trellis section, or equivalently the edge
entropy, and the number of legal vertices for each trellis depth, or
equivalently the vertex entropy, in combination with the interleaver
to improve the algorithm as described below.

Let $v_T$ denote a chosen systematic bit-position within a vertex
entropy transition from level $j$ at time $t$ to level $j+1$ at time
$t+1$ (a forward transition from level $j$ at time $t$), or from level
$j$ at time $t$ to level $j+1$ at time $t-1$ (a backward transition
from level $j$ at time $t$) for one of the constituent trellises.  The
\emph{effective length} of $v_T$ is the number of undetermined
systematic bit-positions that will be determined by the
forward-backward algorithm on the considered constituent trellis
before any interleaving, if $v_T$ is guessed. To simplify notation we
assume that $v_T$ is chosen based on the first constituent trellis.
The effective length of $v_T$ is a positive integer smaller than or
equal to the number of undetermined systematic bit-positions within
the range $[t,t+w_s-1]$ for a forward transition, or within the range
$[t-w_s,t-1]$ for a backward transition. The positive integer $w_s$ is
the smallest integer such that the vertex entropy at time $t+w_s$, for
a forward transition, or at time $t-w_s$, for a backward transition,
is different from $j+1$. A general upper bound on the effective length
of $v_T$ is $w_s$. When $j=0$, we can use the edge entropy to find the
exact value of the effective length of $v_T$. Let $w_e$ denote the
smallest positive integer such that the edge entropy for the
$(t+w_e)$th trellis section, i.e., for the transition from time
$t+w_e$ to time $t+w_e+1$, for a forward transition, or for the
$(t-1-w_e)$th trellis section, for a backward transition, is $2$. A
general upper bound on the effective length of $v_T$ in this case is
$\min(w_s,w_e)$. The exact value is the number of undetermined
systematic bit-positions within the range $[t,t+\min(w_s,w_e)-1]$, for
a forward transition, or within the range $[t-\min(w_s,w_e),t-1]$, for
a backward transition.  The selection of bit-positions can be improved
even further by actively using the interleaver. If we choose a
bit-position $v_T$ with the property that both $v_T$ and $\pi(v_T)$
are within vertex entropy transitions, then the performance will be
improved.

Finally, we remark that the simple version described in
Section~\ref{sec:turbo_improved} provides good results as indicated in
Section~\ref{sec:numerical_sim} below.

\subsection{Some Properties of Improved Turbo Decoding}

In this subsection we establish some basic results of improved turbo
decoding as described above. The following lemma is simple, but
important, since the bit-position selection algorithm in
Section~\ref{sec:turbo_improved} is based on this result.

\begin{lemma} \label{lemma:1}
Apply improved turbo decoding as described above using the
bit-position selection algorithm in Section~\ref{sec:turbo_improved}.
Then, the channel value corresponding to the selected bit-position
$v_T$ is an erasure.
\end{lemma}

\begin{proof}
The bit-position selection algorithm in
Section~\ref{sec:turbo_improved} selects only (systematic)
bit-positions $v_T \in \{0,\dots,I-1\}$ with the property that
$\gamma^{(T)}_{x,\pi_x(v_T)} \neq \gamma^{(T)}_{x,\pi_x(v_T)+1}$ for
$x=a$ or $b$, where $\pi_x(v_T)=v_T$ for $x=a$ and
$\pi_x(v_T)=\pi(v_T)$ for $x=b$. If the channel value is \emph{not}
erased, then we know the $v_T$th information bit with probability $1$.
Since the $v_T$th information bit is known with probability $1$, and
both constituent trellises are information bit-oriented, there is only
a single legal edge out of each legal vertex at trellis depth
$\pi_x(v_T)$ for $x=a$ or $b$. Consequently, the number of legal
trellis vertices at trellis depth $\pi_x(v_T)+1$ is equal to the
number of legal trellis vertices at trellis depth $\pi_x(v_T)$, and
the result follows by contradiction.
\end{proof}

\begin{lemma} \label{lemma10}
Let $\mathcal{C}$ denote a given PCCC with interleaver $\pi$. Let
$\mathcal{S}=\mathcal{S}(\pi)$ denote a turbo stopping set, and erase
all bit-positions in $\mathcal{S}$. Then, choose any bit-position $j
\in \mathcal{S}$, and do the following.
\begin{enumerate}
\item Fix the bit-value in bit-position $j$ to $0$ and perform turbo decoding until either the decoder fails to progress further, or the decoder reaches an inconsistency. If the decoder does not reach an inconsistency, denote the set of erased positions that remain when the decoder stops by $\mathcal{S}_j^{(0)}$.
\item Fix the bit-value in bit-position $j$ to $1$ and perform turbo decoding until either the decoder fails to progress further, or the decoder reaches an inconsistency. If the decoder does not reach an inconsistency, denote the set of erased positions that remain when the decoder stops  by $\mathcal{S}_j^{(1)}$.
\end{enumerate}
If the decoder does \emph{not} reach an inconsistency in either of the two cases above, then the two sets $\mathcal{S}_j^{(0)}$ and $\mathcal{S}_j^{(1)}$ are equal.
\end{lemma}

\begin{proof}
Fix the bit-value in bit-position $j$ to $c$ where $c=0$ or $1$. Then
the number of possible paths in the first constituent code is
immediately reduced by a factor of $2$, since the subcode $\bar{C}_a$
corresponding to $\mathcal{S}$ is linear. Let this reduced set of
legal paths in the first constituent code be denoted by
$P^{(c)}_{a,j}$. Furthermore, the forward-backward algorithm for the
first constituent code will determine additional bit-positions (which
are previously unknown) contained within a set $S_{a,j}^{(c)}$. For a
given bit-position $i \in S_{a,j}^{(c)}$, all paths in $P^{(c)}_{a,j}$
will have the same bit-value of $\tilde{c}$ (depending on the value of
$c$) in this bit-position. Since the subcode $\bar{C}_a$ is linear,
all paths in $P^{(\bar{c})}_{a,j}$, where $\bar{c}$ denotes the
complement of $c$, will also have a fixed bit-value of
$\bar{\tilde{c}}$ in bit-position $i$. Thus, it holds that $i \in
S_{a,j}^{(\bar{c})}$, from which it follows that
$S_{a,j}^{(c)}=S_{a,j}^{(\bar{c})}$, since $c$ and $i$ both are
arbitrarily chosen. For the second constituent code, several
bit-positions are fixed due to extrinsic information from the first
constituent code. However, we can apply the same type of arguments as
above to show that the sequence of additional bit-positions determined
by the forward-backward algorithm in the second constituent code is
independent of $c$. The result follows by applying these arguments in
an iterative fashion until there is no further progress.
\end{proof}

\begin{lemma} \label{lemma11}
Apply improved turbo decoding as described above with the bit-position
selection algorithm in Section~\ref{sec:turbo_improved}. For any two
elements
\begin{displaymath}
L=\{(L_1^{(0)},L_2^{(0)}),\dots,(L_1^{(|L|-1)},L_2^{(|L|-1)})\} \in \mathcal{L} \text{ and } \tilde{L}=\{(\tilde{L}_1^{(0)},\tilde{L}_2^{(0)}),\dots,(\tilde{L}_1^{(|\tilde{L}|-1)},\tilde{L}_2^{(|\tilde{L}|-1)})\} \in \mathcal{L}
\end{displaymath}
with the property that $|L|=|\tilde{L}|$, it holds that
$L_1^{(i)}=\tilde{L}_1^{(i)}$ for all $i$, $i=0,\dots,|L|-1$, i.e.,
the actual bit-values in the guessed bit-positions do not influence on
which bit-positions are selected next by the bit-position selection
algorithm in Section~\ref{sec:turbo_improved}, as long as no
inconsistency is reached.
\end{lemma}

\begin{proof}
The result follows directly from Lemma~\ref{lemma10}.
\end{proof}

We remark that due to Lemma~\ref{lemma11} we can reduce the number of
times we need to run the bit-position selection algorithm from
Section~\ref{sec:turbo_improved} when performing improved turbo
decoding.

\begin{theorem}
Improved turbo decoding is ML decoding on the BEC when $l_{\rm max} \rightarrow \infty$.
\end{theorem} 

\begin{proof}
The proof is two-fold. First we prove that if the algorithm
terminates, then we have an ML decoder. Secondly, we prove that the
algorithm will always terminate.
\begin{itemize}
\item[1)] It follows from the pseudo-code above (step 2, item 2), and step 3, item 5)) that if the algorithm terminates, then the decoder has not reached an inconsistency and $\hat{c}_j \neq \star$ for all $j$ such that $\mu_a(j) \neq \ast$ and $\psi_a(\mu_a(j)) \neq \ast$, $j=0,\dots,N-1$, since $l_{\rm max} \rightarrow \infty$. The original turbo decoding algorithm will not introduce bit errors and neither will the improved turbo decoding algorithm due to items 5) and 7) in step 3. In more detail, the algorithm will not terminate if the decoder has reached an inconsistency, and no further guessing is performed if this is the case. Thus, if the improved turbo decoding algorithm terminates, then transmitted codeword is recovered, or there exists a different turbo codeword $\mathbf{c}'$ with support set $\chi(\mathbf{c}') \subseteq \mathcal{E}$ where $\mathcal{E}$ denotes the subset of remaining erased bit-positions. In the latter case the transmitted codeword is not recovered. An ML decoder will not be able to determine the transmitted codeword in the latter case either, since both codewords $\mathbf{c}$ and $\mathbf{c}+\mathbf{c}'$ where $\mathbf{c}$ denotes the transmitted codeword are equally likely to have been transmitted. Thus, improved turbo decoding is ML decoding.
\item[2)] As the algorithm proceeds, erased bit-positions are guessed. Each time an element $L$ is removed from the list $\mathcal{L}$, turbo decoding is performed. If the decoder does not reach an inconsistency and $\hat{c}_j = \star$ for some $j$ such that $\mu_a(j) \neq \ast$ and $\psi_a(\mu_a(j)) \neq \ast$, $j=0,\dots,N-1$, then a \emph{new} erased bit-position is guessed (see the bit-position selection algorithm in Section~\ref{sec:turbo_improved} and Lemma~\ref{lemma:1} for details). The decoder will always terminate, since there is a finite number of bit-positions to guess.
\end{itemize}
\end{proof}

\subsection{Numerical Results} \label{sec:numerical_sim}
\begin{figure}[htb]
\par
\begin{center}
\includegraphics[width=4.5in,height=4.0in]{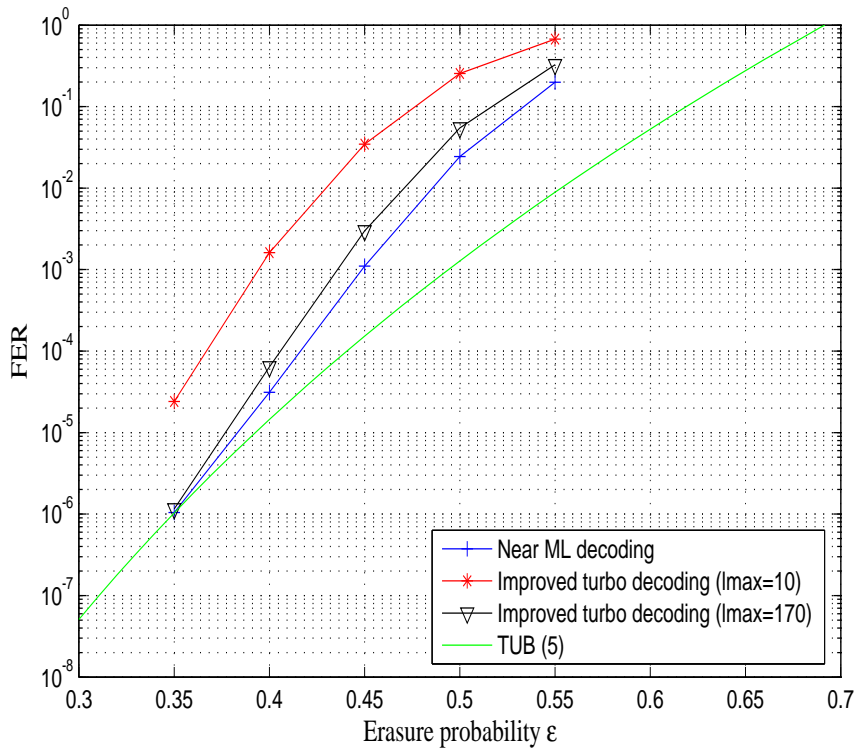}
\end{center}
\caption{\label{figur2} {FER on the BEC of the  $(155,64,18)$ turbo code from Section~\ref{sec:155turbo}.}}
\end{figure}

Here we present some simulation results of improved turbo decoding on
the BEC. The simulated frame error rate (FER) is presented in
Fig.~\ref{figur2} for the $(155,64,18)$ turbo code introduced in
Section~\ref{sec:155turbo}. We have used the bit-position selection
algorithm described in Section~\ref{sec:turbo_improved} in the
simulations. The truncated union bound (TUB) in Fig.~\ref{figur2} is
computed from the first $5$ non-zero terms of the code's weight
distribution.  The near ML decoding curve is obtained using improved
turbo decoding with a large number for $l_{\rm max}$. In
Table~\ref{tab:table3} we have tabulated, for different values of the
channel erasure probability $\epsilon$, the empirical value of $l_{\rm
  max}$ such that improved turbo decoding is near ML decoding. In this
context, improved turbo decoding is near ML decoding when the fraction
of ML-decodable frame errors observed in the simulation is $
\lessapprox 0.05$.  The corresponding estimated values of the expected
number of iterations $E[T]$ are tabulated in the third row of the
table.  The gap between the TUB and the near ML performance curve at
moderate-to-high values of $\epsilon$ is due to the fact that only a
limited number of codewords are taken into account in the summation of
the union bound.  The two remaining curves show the FER for two
different values of $l_{\rm max}$. Observe that when $l_{\rm max}$ is
increased, the performance improves. In Table~\ref{tab:table2}
estimated values of $E[T]$ of improved turbo decoding are tabulated
for different values of $\epsilon$ and $l_{\rm max}$.  From
Table~\ref{tab:table2} we observe that the difference in the estimated
values of $E[T]$ for $l_{\rm max}=170$ and $l_{\rm max}=10$ decreases
when $\epsilon$ decreases. In particular, for $\epsilon=0.40$, there
is almost no difference in the expected number of iterations. However,
as can be seen from Fig.~\ref{figur2}, there is a large difference in
performance. The numbers in Tables~\ref{tab:table3} and
\ref{tab:table2} are based on more than $1000$ observed frame errors
for $\epsilon=0.40, 0.45, 0.50, 0.55$, and more than $100$ frame
errors for $\epsilon=0.35$. Similar performance improvements have been
observed for the $(3600,1194,49)$ turbo code from \cite{cro04}.

\begin{table}[tbp]
\par
\begin{minipage}{\linewidth}
\renewcommand{\thefootnote}%
{\thempfootnote}
\begin{center}
\caption{ {\label{tab:table3} {Estimated expected number of iterations $E[T]$ and empirical $l_{\rm max}$ such that improved turbo decoding is near ML decoding  for the $(155,64,18)$ turbo code from Section~\ref{sec:155turbo}}}}
\begin{tabular}{|c||c|c|c|c|c|} \hline
$\epsilon$ & $0.55$ & $0.50$ & $0.45$ & $0.40$ & $0.35$ \\ \hline \hline
$l_{\rm max}$ & $2600$ & $2400$ & $2000$ & $1100$ & $400\footnote{The estimate is less reliable, since only $6$ ML-decodable frame errors have been observed compared to about $50$ for $\epsilon > 0.35$.}$ \\ \hline
$E[T]$ & $402$ & $49$ & $4.4$ & $1.46$ & $1.10$ \\ \hline
\end{tabular}%
\end{center}%
\end{minipage}%
\end{table}
\begin{table}[tbp]
\par
\begin{center}
\caption{ \label{tab:table2} {Estimated expected number of iterations $E[T]$ of improved turbo decoding for different values of $\epsilon$ and  $l_{\rm max}$ for the $(155,64,18)$ turbo code from Section~\ref{sec:155turbo}}}
\begin{tabular}{|c||c|c|c|c|c|} \hline
$l_{\rm max} \downarrow$, $\epsilon \rightarrow$ & $0.55$ & $0.50$ & $0.45$ & $0.40$ & $0.35$ \\ \hline \hline
$10$ & $8.3$ & $5.1$ & $2.42$ & $1.42$ & $1.10$ \\ \hline
$170$ & $74$ & $19.5$ & $3.57$ & $1.45$ & $1.10$ \\ \hline
\end{tabular}
\end{center}
\end{table}

\section{Turbo Stopping Set Enumeration} \label{sec:alg}

\subsection{Convolutional Codes and Trellises}

An edge-labeled directed graph is a triple $(V,E,A)$, consisting of a set $V$ of vertices, a finite set $A$ called the alphabet, and a set $E$ of ordered triples $(v,a,v')$, with $v,v' \in V$ and $a \in A$ called edges. The edge $(v,a,v')$ begins at $v$, ends at $v'$, and has label $a$.

Let $C$ denote a linear $(n,k,\nu)$ convolutional code over some
finite field $F_q$ of $q$ elements, where $\nu$ is the constraint
length or the code degree. In this work the convolutional code symbols
are taken from the binary field $F_2=GF(2)$. A convolutional code can
be defined by an $(n-k) \times n$ polynomial parity-check matrix
${\mathbf{H}}(D)$. We assume in general a \emph{canonical}
parity-check matrix \cite{Mce98}.  The maximum degree of the
polynomials in the $i$th row is the $i$th row degree, denoted by
$\nu_i$.

Let ${\mathbf{H}}^{L}$ be  the matrix ${\mathbf{H}}(0)$, and let ${\mathbf{H}}^{H}$ be  the matrix $\diag(D^{\nu_1},\dots,D^{\nu_{n-k}}) {\mathbf{H}}(D^{-1})$ with $D=0$. When a matrix is given an integer interval subscript we mean the submatrix consisting of the columns with indices in the interval. The columns in a matrix are indexed from left to right with positive integers.

\subsubsection{Trellis Representation of Convolutional Codes} \label{sec:conv}
The \emph{minimal} trellis of $C$ can be constructed from a
parity-check matrix of the code as outlined in \cite{sid94}.  The
minimal trellis can be regarded (after an initial transient) as the
infinite composition of a basic building block which is called the
\emph{trellis module}. The trellis module $T=(V,E,F_q)$ of $C$ is an
edge-labeled directed graph with the property that the vertex set $V$
can be partitioned as
\begin{equation} \label{eq:partition}
V = V_0 \cup V_1 \cup \cdots \cup V_n
\end{equation}
such that every edge in $E$ begins at a vertex in $V_i$ and ends at a
vertex in $V_{i+1}$, for some $i$, $i=0,\dots,n-1$. The \emph{depth}
of the trellis module is $n$. The ordered index set
$\mathcal{I}=\{0,1,\dots,n\}$ induced by the partition in
(\ref{eq:partition}) is called the \emph{time axis} for $T$. The
partition in (\ref{eq:partition}) also induces a partition $E = E_0
\cup E_1 \cup \cdots \cup E_{n-1}$ of the edge set $E$ where $E_i$ is
the subset of edges that begin at a vertex in $V_i$ and end at a
vertex in $V_{i+1}$.

Define $b_0=0$ and $b_i = \rank[{\mathbf{H}}_{n-i+1,n}^{L}]$, $i
=1,\dots,n$, and $f_0=0$ and $f_i= \rank[{\mathbf{H}}_{1,i}^{H}]$, $i
=1,\dots,n$.  The vertex set $V_i$ is a vector space of dimension
$\dim(V_i) = \nu-n+k+f_i+b_{n-i} \leq \nu+n-k$, from which it follows
that $\dim(V_0)=\dim(V_n)=\nu$. The edge set $E_i$ is also a vector
space of dimension $\dim(E_i) = \nu-n+k+f_i+b_{n-i-1}+1$.

Let $\mathcal{I}_{\rm info}$ be the subset of $\mathcal{I} \setminus
\{n\}$ consisting of \emph{all} integers $i$ with the property that
$b_{n-i}=b_{n-i-1}$. Furthermore, we assume without loss of generality
that $b_n=b_{n-1}$ which implies that $0 \in \mathcal{I}_{\rm info}$.
Let $n_i=j+1$ where $j$ is the largest non-negative integer $\leq
n-i-1$ such that $b_{n-i-1} \neq \cdots \neq b_{n-i-j-1}$ for every $i
\in \mathcal{I}_{\rm info}$.  As argued, for instance in \cite{cha96},
there are $n-k$ time instances $i \in \mathcal{I} \setminus
\mathcal{I}_{\rm info}$ in which there is only a single edge out of
each vertex in $V_i$. By \emph{sectionalization} the depth of the
trellis module $T$ can be reduced to $k$. This reduced trellis module
is called an information bit-oriented trellis module and is denoted by
$T_{\rm info}=(V_{\rm info},E_{\rm info},\cup_{i \in \mathcal{I}_{\rm
    info}} F_q^{n_i})$ where $V_{\rm info}= \cup_{i \in
  \mathcal{I}_{\rm info} \cup \{n\}} V_i$ and $E_{\rm info}= \cup_{i
  \in \mathcal{I}_{\rm info}} E'_i$. The edge set $E'_i$ is the set of
paths that begin at a vertex in $V_i$ and end at a vertex in
$V_{i+n_i}$. The label of an edge in $E'_i$ is the label sequence
along the defining path which is a $q$-ary sequence of length $n_i$.
The edge set $E'_i$ is a vector space of the same dimension as $E_i$.

In the trellis module $T_{\rm info}$ there are $q$ edges out of each
vertex in $V_{\rm info} \setminus V_n$.  Thus, we can assign a $q$-ary
input label to each edge $e \in E_{\rm info}$, and the trellis module
$T_{\rm info}$ can used for encoding.

\subsubsection{Trellis Representation of Subcodes of Convolutional Codes}
A trellis representing subcodes of $C$ can be written (after an
initial transient) as the infinite composition of an \emph{extended}
trellis module $\bar{T}=(\bar{V},\bar{E},F_2)$ where $\bar{V}=
\bar{V}_0 \cup \cdots \cup \bar{V}_n$ and $\bar{E}= \bar{E}_0 \cup
\cdots \cup \bar{E}_{n-1}$ are partitions of $\bar{V}$ and $\bar{E}$,
respectively. Each vertex of $\bar{V}_i$ corresponds to a subspace of
$V_i$, and each edge of $\bar{E}_i$ corresponds to a subspace of
$E_i$. The number of distinct $k$-dimensional subspaces of an
$n$-dimensional vector space over $F_q$, $k =1,\dots,n$, denoted by
$S(k,n,q)$, is \cite[p.\ 444]{MCWSL}
\begin{displaymath}
S(k,n,q) = \frac{(q^n-1)(q^{n-1}-1) \cdots (q^{n-k+1}-1)}{(q^k-1)(q^{k-1}-1) \cdots (q-1)} = \left[  \genfrac{}{}{0pt}{}{n}{k} \right]_q
\end{displaymath}
from which it follows that
\begin{displaymath}
|\bar{V}_i| = 1+\sum_{j=1}^{\dim(V_i)} \left[ \genfrac{}{}{0pt}{}{\dim(V_i)}{j} \right]_q\;\text{and}\; |\bar{E}_i| = 1+\sum_{j=1}^{\dim(E_i)} \left[ \genfrac{}{}{0pt}{}{\dim(E_i)}{j} \right]_q.
\end{displaymath}

The left (resp.\ right) vertex of an edge $e$ is denoted by $v^L(e)$
(resp.\ $v^R(e)$). The label of an edge $e$ is denoted by $c(e)$. Note
that the edge $e$ could either be an edge in the trellis module $T$ or
in the extended trellis module $\bar{T}$.

The connections in the extended trellis module $\bar{T}$ are
established as follows. An edge $\bar{e} \in \bar{E}_i$ corresponds to
a subspace of $E_i$ of dimension $d(\bar{e})$ and basis
$\{e_0(\bar{e}),\dots,e_{d(\bar{e})-1}(\bar{e}) \}$.  The vertex in
$\bar{V}_i$ that corresponds to the vector space spanned by $\{
v^L(e_0(\bar{e})),\dots,v^L(e_{d(\bar{e})-1}(\bar{e})) \}$ is
connected to the vertex in $\bar{V}_{i+1}$ that corresponds to the
vector space spanned by $\{
v^R(e_0(\bar{e})),\dots,v^R(e_{d(\bar{e})-1}(\bar{e})) \}$ by the edge
$\bar{e}$. The binary label $c(\bar{e})$ of $\bar{e}$ is $1$ if at
least one of the $q$-ary labels $c(e_j(\bar{e}))$ of $e_j(\bar{e}) \in
E_i$, $j=0,\dots,d(\bar{e})-1$, is non-zero. Otherwise, it is $0$. In
the case of parallel edges in $\bar{T}$ with the same label, we keep
only one.

Note that an information bit-oriented extended trellis module
$\bar{T}_{\rm info}$ can be obtained by sectionalization as described
above.  A trellis $\bar{\mathcal{T}}_{\rm info}$ constructed as the
infinite composition of the trellis module $\bar{T}_{\rm info}$ has
paths that are in one-to-many correspondence with subcodes of $C$. The
label sequence of a path in $\bar{\mathcal{T}}_{\rm info}$ is a binary
sequence where the set of $1$-positions is equal to the support set of
the subcodes represented by the given path. Since distinct subcodes
could have equal support sets, there could be paths in
$\bar{\mathcal{T}}_{\rm info}$ that have equal label sequences.

Input labels can be assigned to the edges in the trellis module
$\bar{T}_{\rm info}$, and thus $\bar{T}_{\rm info}$ can be used for
\emph{encoding}, but the encoding is \emph{not} one-to-one, since
there could be more than one path in $\bar{\mathcal{T}}_{\rm info}$
with the same input (and output) label sequence.

\begin{example} \label{ex:1}
Consider the $(2,1,2)$ binary convolutional code defined by the parity-check matrix $\mathbf{H}(D) = ( 1+D^2\;1+D+D^2)$.
The trellis module $T_{\rm info}$ and the extended trellis module $\bar{T}_{\rm info}$ are both  depicted in Fig.~\ref{trellis}. Note that the trellis module $\bar{T}_{\rm info}$ is \emph{non-linear}.

\begin{figure}[htb]
\par
\begin{center}
\includegraphics[width=3.5in,height=2.5in]{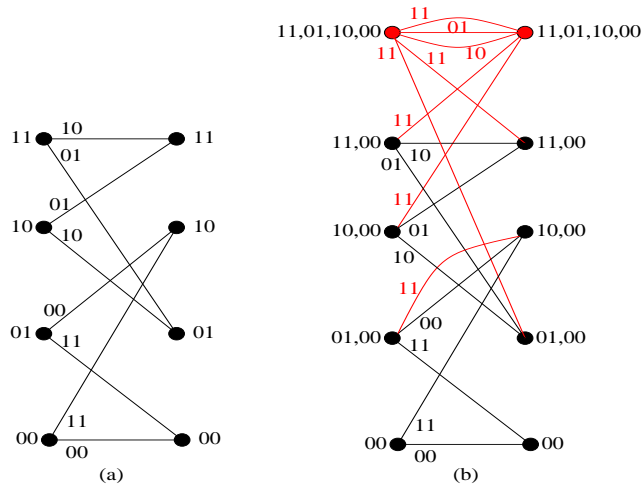}
\end{center}
\caption{\label{trellis} {(a) Basic trellis module $T_{\rm info}$, and (b) extended trellis module $\bar{T}_{\rm info}$ of the convolutional code from Example~\ref{ex:1}.}}
 \end{figure} 
\end{example}

The numbers of vertices and edges of an (extended) information
bit-oriented trellis module $T_{\rm info}$ divided by $k$ are called
the vertex and edge complexities of $T_{\rm info}$, and are given,
respectively, by
\begin{displaymath}
\mu(T_{\rm info})=\frac{1}{k}\sum_{i \in \mathcal{I}_{\rm info}} |V_i| \text{ and } \phi(T_{\rm info})=\frac{1}{k}\sum_{i \in \mathcal{I}_{\rm info}} |E_i|.
\end{displaymath}
The vertex and edge complexities of $T_{\rm info}$ and $\bar{T}_{\rm
  info}$ are tabulated in Table~\ref{table1} for different values of
$\nu$ for $(n,1,\nu)$ binary convolutional codes.
\begin{table}[tbp] 
\par
\begin{center}
\caption{\label{table1} {Vertex and edge complexities of the trellis modules $T_{\rm info}$ and $\bar{T}_{\rm info}$ for  $(n,1,\nu)$ binary convolutional codes}}
\begin{tabular}{|c||c|c|c|c|c|c|} \hline
$\nu$  \T \B & 2 & 3 & 4 & 5 & 6  \\ \hline \hline
$\mu(T_{\rm info})$ \T \B & 4 & 8 & 16 & 32 & 64 \\ \hline
$\phi(T_{\rm info})$ \T \B & 8 & 16 & 32 & 64 & 128 \\ \hline 
$\mu(\bar{T}_{\rm info})$ \T \B & 5 & 16 & 67 & 374 & 2825 \\ \hline
$\phi(\bar{T}_{\rm info})$ \T \B & 16 & 67 & 374 & 2825 & 29212 \\ \hline
\end{tabular}
\end{center}
\end{table}

\subsection{The Uniform Interleaver}
Consider an $(n,k,\nu)$ systematic convolutional code which is
terminated to the all-zero vertex at trellis depth $I \geq \lceil
\nu/k \rceil k$ in $\mathcal{T}_{\rm info}$ where $I$ is assumed to be
a multiple of $k$. The resulting linear block code $C$ has length
$\delta=(I/k)n$ and dimension $I- \nu$.  When $\nu=0$, the
convolutional code is actually a linear block code in which case we
choose $I=k$.

Partition all the subcodes of $C$ of dimension $d$, $d=1,\dots,I-\nu$,
into equivalence classes based on their support sets. In particular,
all subcodes within a specific subcode class are required to have the
same support set, but the subcodes may have different dimensions.

We define the \emph{subcode input-redundancy support size enumerating
  function} (SIRSEF) of $C$. The SIRSEF has the form
\begin{equation}
A^{C}(W,Z)=\sum_{w=1}^{I} \sum_{z=0}^{\delta-I} a^{C}_{w,z} W^wZ^z
\end{equation}
where $a^{C}_{w,z}$ is the number of subcode classes of $C$ of \emph{input} support set size $w$ and \emph{parity} support set size $z$.
When analyzing the performance it is useful to group the terms in the SIRSEF according to input support set size. The \emph{conditional} SIRSEF
\begin{equation}
A^{C}_w(Z) = \sum_{z=0}^{{\delta}-I} a^{C}_{w,z}Z^z
\end{equation}
 enumerates subcode classes of different parity support set sizes associated with a particular input support set size. 
The conditional SIRSEF $A^{C}_w(Z)$ and the SIRSEF $A^{C}(W,Z)$ are related to one another through the following pair of expressions
\begin{equation} \label{eq:sirsef1}
A^{C}(W,Z) = \sum_{w=1}^{I} W^wA^{C}_w(Z)  \; {\rm and}\;
A^{C}_w(Z) = \frac{1}{w\,!} \cdot \frac{\partial^w A^{C}(W,Z)}{\partial W^w} \Big\arrowvert_{W=0}.
\end{equation}

Let $A_w^{C_x}(Z)$, $x=a,b$, denote the conditional SIRSEF for the
constituent code $C_x$ of a given PCCC $\mathcal{C}$ with interleaver
length $I$. We assume that the interleaver is uniform. With a uniform
interleaver, the $\binom{I}{w}$ possible sequences of length $I$ and
weight $w$ occur with equal probability at the output of the
interleaver when the interleaver is fed with a length-$I$ and
weight-$w$ sequence. Let $S^{\mathcal{C}}(W,Z)$ denote the
\emph{input-redundancy turbo stopping set size enumerating function}
(IRTSSEF) of $\mathcal{C}$. The IRTSSEF has the form
\begin{equation}
S^{\mathcal{C}}(W,Z) = \sum_{w=1}^{I} \sum_{z=0}^{2(\delta-I)} s_{w,z}^{\mathcal{C}} W^wZ^z
\end{equation}
where $s_{w,z}^{\mathcal{C}}$ is the number of turbo stopping sets of $\mathcal{C}$ of \emph{input} size $w$ and \emph{parity} size $z$.
 The conditional IRTSSEF 
\begin{equation}
S_w^{\mathcal{C}}(Z) = \sum_{z=0}^{2(\delta-I)} s_{w,z}^{\mathcal{C}} Z^z
\end{equation}
enumerates turbo stopping sets of
 different parity sizes associated with a particular input size.
The conditional IRTSSEF $S_w^{\mathcal{C}}(Z)$ and the IRTSSEF $S^{\mathcal{C}}(W,Z)$ are related to one another through the following pair of expressions
\begin{equation} \label{eq:dirtsef}
S^{\mathcal{C}}(W,Z) = \sum_{w=1}^{I} W^wS^{\mathcal{C}}_w(Z) \; {\rm and}\;
S^{\mathcal{C}}_w(Z) = \frac{1}{w\,!} \cdot \frac{\partial^w S^{\mathcal{C}}(W,Z)}{\partial W^w} \Big\arrowvert_{W=0}.
\end{equation}
A PCCC with a uniform interleaver has a uniform probability of
matching a given support set in $A_w^{C_a}(Z)$ with any given support
set in $A_w^{C_b}(Z)$. Thus, it follows that the conditional IRTSSEF
for $\mathcal{C}$ is
\begin{equation} \label{eq:stopping1}
S_w^{\mathcal{C}}(Z) = \frac{A_w^{C_a}(Z)A_w^{C_b}(Z)}{\binom{I}{w}}.
\end{equation}
The \emph{turbo stopping set size enumerating function} (TSSEF) for $\mathcal{C}$ is
\begin{equation} \label{eq:tssef}
S^{\mathcal{C}}(X)=\sum_{i=1}^{2\delta-I} s_i^{\mathcal{C}} X^i,\;\text{where}\; s^{\mathcal{C}}_i=\sum_{w=1}^{\min(i,I)} s^{\mathcal{C}}_{w,i-w}
\end{equation}
and $s^{\mathcal{C}}_i$ 
is the number of turbo stopping sets of size $i$. If only $1$-dimensional subcodes are considered in the constituent conditional SIRSEFs $A_w^{C_a}(Z)$ and $A_w^{C_b}(Z)$, then we get the weight enumerating function (WEF) for $\mathcal{C}$.

\subsubsection{The $(7,4)$ Hamming Code} \label{sec:hamming}

We consider the $(7,4)$ Hamming code in its cyclic form in which case $\nu=0$, $I=k=4$, and $\delta=n=7$. The SIRSEF of the Hamming code is
\begin{displaymath} \label{eq:Hamming}
W(3Z^2+Z^3)+W^2(3Z+3Z^2+6Z^3)+
W^3(1+3Z+12Z^2+4Z^3)
+W^4(3Z+3Z^2+Z^3).
\end{displaymath}
The IRTSSEF for the PCCC $\mathcal{C}$ with a uniform interleaver is
\begin{equation} \label{eq:Hamming1}
\begin{split}
&W(2.25Z^4+1.5Z^5+0.25Z^6)+W^2(1.5Z^2+3Z^3+7.5Z^4+6Z^5+6Z^6)+\\
&W^3(0.25+1.5Z+8.25Z^2+20Z^3
+42Z^4+24Z^5+4Z^6)+W^4(9Z^2+18Z^3+15Z^4+6Z^5+Z^6)
\end{split}
\end{equation}
from which the TSSEF can be calculated. The result is
\begin{displaymath}
0.25X^3+3X^4+13.5X^5+38X^6+66.25X^7+45X^8+10X^9+X^{10}.
\end{displaymath}
Note that the WEF for ${\mathcal{C}}$ is
\begin{displaymath}
1+0.25X^3+3X^4+7.5X^5+3X^6+0.25X^7+X^{10}.
\end{displaymath}
We can check the result in (\ref{eq:Hamming1})  by computing the IRTSSEFs of the PCCCs constructed using all the $4\,! = 24$ possible interleavers. The results are tabulated in Table~\ref{table:Hamming}.
\begin{table}[tbp]
\par
\begin{center}
\caption{\label{table:Hamming} {IRTSSEFs, TSSEFs, and WEFs of the PCCCs constructed using all the $4\,! = 24$ possible interleavers and with the $(7,4)$ Hamming  code as constituent codes.}}
\begin{tabular}{|c|c|} \hline
$\pi$ & $S^{\mathcal{C}}(W,Z)$, $S^{\mathcal{C}}(X)$, and  ${\rm WEF}(X)$ \\ \hline
3210 &  \multirow{18}{*}{$\begin{array}{rcl} S^{\mathcal{C}}(W,Z) & = & 1+W(2Z^4+2Z^5)\\
&+ & W^2(Z^2+4Z^3+7Z^4+6Z^5+6Z^6) \\
&+ & W^3(2Z+8Z^2+20Z^3+42Z^4+24Z^5+4Z^6)\\
&+ & W^4(9Z^2+18Z^3+15Z^4+6Z^5+Z^6) \\
S^{\mathcal{C}}(X) & = & 1+3X^4+14X^5+38X^6+66X^7+45X^8+10X^9+X^{10} \\
{\rm WEF}(X) & = & 1+3X^4+8X^5+3X^6+X^{10} \end{array}$} \\
3201 & \\
3120 & \\
3102 & \\
3012 & \\
3021 & \\
2310 & \\
2301 & \\
2130 & \\
2031 & \\
1230 & \\
1320 & \\
1302 & \\
1032 & \\
0231 & \\
0132 & \\
0312 & \\
0321 & \\ \hline
2103 &  \multirow{6}{*}{$\begin{array}{rcl} S^{\mathcal{C}}(W,Z) & = & 1+W(3Z^4+Z^6)+W^2(3Z^2+9Z^4+6Z^5+6Z^6)\\
& + & W^3(1+9Z^2+20Z^3+42Z^4+24Z^5+4Z^6) \\
& + & W^4(9Z^2+18Z^3+15Z^4+6Z^5+Z^6) \\
S^{\mathcal{C}}(X) &= & 1+X^3+3X^4+12X^5+38X^6+67X^7+45X^8+10X^9+X^{10} \\
{\rm WEF}(X) & =& 1+X^3+3X^4+6X^5+3X^6+X^7+X^{10} \end{array}$} \\
2013 &\\
1203 & \\
1023 & \\
0213 & \\
0123 & \\ \hline
\end{tabular}
\end{center}
\end{table}
Only two types of IRTSSEF are possible. It is easy to verify that the
average over all possible interleavers is equal to the expression in
(\ref{eq:Hamming1}). Note that the WEF is dependent on the interleaver
while the \emph{non-codeword} TSSEF is the same for \emph{all}
interleavers.

\subsubsection{Convolutional Codes as Constituent Codes}

Let $T^{C_x}(W,Z,\Gamma,\Sigma)$, $x=a,b$, enumerate all subcode
classes of $C_x$ constructed from trellis paths in constituent trellis
$\bar{\mathcal{T}}^x_{\rm info}$ leaving the all-zero vertex at
trellis depth zero, and remerging into the all-zero vertex at or
before trellis depth $I$, with possible remerging into the all-zero
vertex at other depths in between, subject to the constraint that,
after remerging, the paths leave the all-zero vertex at the same
trellis depth. In general,
\begin{equation}
T^{C_x}(W,Z,\Gamma,\Sigma) =  \sum_{w=1}^I \sum_{z=0}^{\delta-I} \sum_{\gamma=2}^{I} \sum_{\sigma=1}^{\lfloor \gamma/2 \rfloor} t^{C_x}_{w,z,\gamma,\sigma} W^wZ^z\Gamma^{\gamma}\Sigma^{\sigma}
\end{equation}
where $t^{C_x}_{w,z,\gamma,\sigma}$ is the number of subcode classes
of $C_x$ of input support set size $w$ and parity support set size $z$
constructed from trellis paths of length $\gamma$, and with $\sigma$
remergings with the all-zero vertex. Notice that each subcode class in
$T^{C_x}(W,Z,\Gamma,\Sigma)$ of input support set size $w$ and parity
support set size $z$ constructed from trellis paths of length
$\gamma$, and with $\sigma$ remergings with the all-zero vertex, gives
rise to $\binom{I-\gamma+\sigma}{\sigma}$ subcode classes with the
same input and parity support set sizes. Thus, the conditional SIRSEF
$A_w^{C_x}(Z)$ can be written as
\begin{equation} \label{eq:transfer}
A_w^{C_x}(Z) =   \sum_{z=0}^{\delta-I} \left[ \sum_{\gamma=2}^{I} \sum_{\sigma=1}^{\lfloor \gamma/2 \rfloor} \binom{I-\gamma+\sigma}{\sigma} t^{C_x}_{w,z,\gamma,\sigma} \right] Z^z.
\end{equation}
Finding a closed-form expression for the conditional SIRSEF (as we did
in Section~\ref{sec:hamming} when the constituent codes were Hamming
codes) for a given (large) value of the interleaver length $I$ is
difficult. For this reason, we will use an algorithmic approach to
compute the most significant terms of the conditional SIRSEF in
(\ref{eq:transfer}). One approach is to use the algorithm to be
described in Section~\ref{GPB} with only one constituent code. See
Section~\ref{GPB} below for details.

\subsection{Enumeration of Small-Size Turbo Stopping Sets for a Particular Interleaver} \label{GPB}

Let $T_{\rm info}^x$ and $\bar{T}_{\rm info}^x$ denote the information
bit-oriented and the extended information bit-oriented trellis modules
of constituent code $C_x$, $x=a,b$. We assume that $T_{\rm info}^x$
and $\bar{T}_{\rm info}^x$ both have input and output labels assigned
to the edges. The trellises constructed from the trellis modules
$T_{\rm info}^x$ and $\bar{T}_{\rm info}^x$ are denoted by
$\mathcal{T}_{\rm info}^x$ and $\bar{\mathcal{T}}_{\rm info}^x$,
respectively.

Let $\mathcal{S}_{C_x}$ denote the subset of $\{0,1\}^{N_x}$ of label
sequences of paths that begin and end at the all-zero vertex at
trellis depths $0$ and $I$ of the trellis $\bar{\mathcal{T}}_{\rm
  info}^x$, $x=a,b$. Finally, let
$\mathcal{S}_{\mathcal{C}}=\mathcal{S}_{\mathcal{C}}(K,\mathcal{S}_{C_a},\mathcal{S}_{C_b},\pi)$
denote the PCCC with information length $K$, constituent encoders
$\mathcal{S}_{C_a}$ and $\mathcal{S}_{C_b}$, and interleaver $\pi$.
Note that the support sets of words of $\mathcal{S}_{\mathcal{C}}$ are
in one-to-one correspondence with \emph{all} turbo stopping sets of
$\mathcal{C}$.  Thus, the minimum Hamming weight of $\mathcal{S}_{\cal
  C}$ is equal to the minimum turbo stopping set size of ${\cal C}$.
In general, the set of turbo stopping sets can be obtained by
\emph{turbo encoding} using the constituent encoders
$\mathcal{S}_{C_a}$ and $\mathcal{S}_{C_b}$. Note that there could
exist several paths in $\bar{\mathcal{T}}_{\rm info}^x$ with the same
input label sequence from which it follows that the complexity of
turbo encoding could be more than linear in $K$.

Let $\Pi(\mathcal{S}_{\cal C},\tau)$ be the
problem of finding all words  of ${\mathcal{S}_{\cal C}}$ of Hamming weight
$\leq \tau$.  This problem is equivalent of finding an exhaustive list of all turbo stopping sets of size $\leq \tau$. To simplify notation we assume below that the $I-K$ redundant systematic bits appear at the end of the input block.

A \emph{constraint set} $F$ is a set $ \{(p_i,u_{p_i}) : u_{p_i} \in
\{0,1\}\;\forall p_i \in \Gamma_{p} \}$, where
$\Gamma_{p}\subseteq\{0,\ldots,K-1\}$ is a set of distinct positions.
For any constraint set $F$, let $U^{(F)}$ be the set of length-$K$
vectors $\{{\bf u}=(u_0,\ldots,u_{K-1}) : u_j=u \mbox{ if }(j,u) \in
F, u_j \in \{ 0,1 \} \mbox{ if } j \not \in \Gamma_{p} \}$. Let the
\emph{length} $l=l(F)$ be the number of constraints. We will start
with a constraint set $F$ of length $l$ of the form
$\{(0,u_0),(1,u_1),\ldots,(l-1,u_{l-1})\}$, i.e., it applies
consecutively to the first $l$ positions. When the turbo interleaver
$\pi$ acts on $F$, we obtain a new constraint set $\pi F =
\{(\pi(p_i),u_{p_i})\}$, where in general the constrained positions
are scattered over the input block.

Let ${\mathcal{S}_{\cal C}^{(F)}}$ be the subset of $\mathcal{S}_{\cal
  C}$ that is obtained by encoding the input vectors in $U^{(F)}$, let
$w(F)$ be the minimum Hamming weight of ${\mathcal{S}_{\cal
    C}^{(F)}}$, and let $w^{\prime}(F)$ be any lower bound for $w(F)$.
The pseudo-code of the algorithm to solve $\Pi(\mathcal{S}_{\cal
  C},\tau)$ is given below. Note that the algorithm has the same
\emph{structure} as the algorithm proposed by Garello \emph{et al.} in
\cite{gar01} to solve $\Pi({\cal C},\tau)$. We will refer to this
algorithm as the GPB algorithm. However, there are some differences
that we will discuss below.

\vspace{0.25cm}
\begin{ttfamily} \fontsize{8}{9}\selectfont {
\begin{tabbing}
$/*$ Find all words of $\mathcal{S}_{\cal C}$ of Hamming weight $\leq \tau$ $*/$\\
Add an empty constraint set $F$ to a\\pre\=viously empty list $L$ of
constraint sets.\\ $(\dagger)$  \> {\bf If}  $L$ \= is \= empty\=, terminate the process.
\\ \>{\bf Otherwise}, \\
\>\>choose and take out a constraint set $F$ from $L$.\\
\>{\bf If} $w'(F) \leq \tau$, then\\
   \> \> {\bf If} \= the length $l$ of $F$ is equal to $K$ then:\\
   \> \>\> The single vector in $U^{(F)}$ produces\\
   \> \>\> low-weight words in ${\mathcal{S}_{\cal C}}$ which are saved.\\
   \> \> {\bf Otherwise},\\
    \>\>\> construct two new constraint sets:\\
  \> \>\>\> $F^{\prime} = F \cup \{(l,0)\}$ and $F^{\prime \prime} = F \cup \{(l,1)\}$.\\
 \> \> \> Add  $F^{\prime}$ and $F^{\prime \prime}$ to $L$.\\
\>Proceed from $(\dagger)$.
\end{tabbing}}
\end{ttfamily}

Let ${\mathcal{S}}_{{C}_a}^{(F)}$ be the subset of words generated by
the constituent encoder ${\mathcal{S}}_{C_a}$ when the input vectors
are contained in $U^{(F)}$, and let $w_a(F)$ be the minimum Hamming
weight of ${\mathcal{S}}_{{C}_a}^{(F)}$. Select any vector from
$U^{(F)}$ as an input sequence of ${\mathcal{S}}_{C_a}$. After $l(F)$
time units the encoder has reached a subset $\{ {\sigma}_{a,0}^{(F)},
\dots,{\sigma}_{a,\rho(F)-1}^{(F)} \}$ of cardinality $\rho(F)$ of the
set of trellis vertices of $\bar{\mathcal{T}}_{\rm info}^a$ at trellis
depth $l(F)$. The trellis path of $\bar{\mathcal{T}}_{\rm info}^a$
from the all-zero vertex at trellis depth $0$ to vertex
$\sigma_{a,i}^{(F)}$ at trellis depth $l(F)$ of minimum Hamming weight
is denoted by ${\bf c}_{a,i}^{(F)}$.  Let $w({\bf c}_{a,i}^{(F)})$ be
the Hamming weight of ${\bf c}_{a,i}^{(F)}$, and let
$w(\sigma_{a,i}^{(F)},l(F))$ be the minimum Hamming weight of any path
from vertex $\sigma_{a,i}^{(F)}$ at trellis depth $l(F)$ to the
all-zero vertex at trellis depth $I$. In general, it holds that
\begin{equation} 
\label{wca_eq} w_{a}(F) = \min_{i=0,\dots,\rho(F)-1} \left( w({\bf
c}_{a,i}^{(F)})+w(\sigma_{a,i}^{(F)},l(F)) \right).
\end{equation}
The weights $w(\sigma_{a,i}^{(F)},l(F))$ can be computed in a
preprocessing stage using the Viterbi algorithm. Actually, the weights
$w(\sigma_{a,i}^{(F)},l(F))$ depend only on the vertex
$\sigma_{a,i}^{(F)}$ if $l(F)$ is not too close to $K$ and
$\bar{\mathcal{T}}_{\rm info}^a$ is \emph{non-catastrophic}. This
reduces the memory requirements. However, the weights $w({\bf
  c}_{a,i}^{(F)})$ have to be computed during the course of the
algorithm. In general, the weights $w({\bf c}_{a,i}^{(F)})$ can be
computed by a \emph{constrained} Viterbi algorithm. Note that in the
original GPB algorithm there is no need to apply a constrained Viterbi
algorithm here, since $\rho(F)=1$, and there is a unique path in
$\mathcal{T}_{\rm info}^a$ from the all-zero vertex at trellis depth
$0$ to the vertex $\sigma_{a,0}^{(F)}$ at trellis depth $l(F)$.

Similarly, let $\mathcal{S}_{{C}_b}^{(\pi F)}$ be the subset of words
generated by the constituent encoder $\mathcal{S}_{C_b}$ when the
input vectors are contained in $U^{(\pi F)}$. Also, let $w_b(\pi F)$
be the minimum Hamming \emph{parity} weight of
$\mathcal{S}_{{C}_b}^{(\pi F)}$. We have
\begin{equation}
\label{wcbound_eq}
  w_{\rm bound}(F) = w_a(F)+w_b(\pi F) \leq w(F).
\end{equation}
Note that $w_{\rm bound}(F)$ is a lower bound on $w(F)$, since the
sequence of input bits giving the minimum-weight path in the second
constituent encoder trellis $\bar{\mathcal{T}}_{\rm info}^b$ is not
necessarily an interleaved version of the sequence of input bits
giving the minimum-weight path in the first constituent encoder
trellis $\bar{\mathcal{T}}_{\rm info}^a$. The value of $w_b(\pi F)$
can be determined by the use of a constrained Viterbi algorithm. Since
the positions of $\pi F$ are in general not consecutive, the
complexity of calculating the value of $w_b(\pi F)$ by a constrained
Viterbi algorithm is larger than the complexity of calculating the
weights $w(\mathbf{c}_{a,i}^{(F)})$, $i=0,\dots,\rho(F)-1$, needed in
(\ref{wca_eq}).

In \cite{ros03_12} we outlined several improvements to the basic GPB
algorithm for solving $\Pi(\cal C,\tau)$. All of the improvements
described in the context of solving $\Pi(\cal C,\tau)$ can be applied
when solving $\Pi(\mathcal{S}_{\cal C},\tau)$.

From Table~\ref{table1} we observe that the edge complexity of
$\bar{T}_{\rm info}$ is large compared to the edge complexity of
$T_{\rm info}$ even for $\nu=4$. To reduce complexity we propose to
remove some of the edges from $\bar{T}_{\rm info}^x$. For instance,
all edges in $\bar{T}_{\rm info}^x$ that correspond to edge subspaces
of $T_{\rm info}^x$ of dimension $\geq \alpha$, for some integer
$\alpha \geq 2$, can be removed.

\subsection{Remarks}

We remark that in principle every \emph{trellis-based} turbo code
weight spectrum computation or estimation algorithm can be adapted in
a straightforward manner to find small-size turbo stopping sets. The
only requirement is that the basic trellis module is substituted with
the extended trellis module introduced above. We have considered the
impulse methods by Berrou and Vaton \cite{ber02}, Vila-Casado and
Garello \cite{vil04}, and Crozier \emph{et al.} \cite{cro04} with
promising results.

\subsection{Numerical Results} \label{sec:numerical1}

\begin{figure}[htb]
\par
\begin{center}
\includegraphics[width=4.5in,height=4.0in]{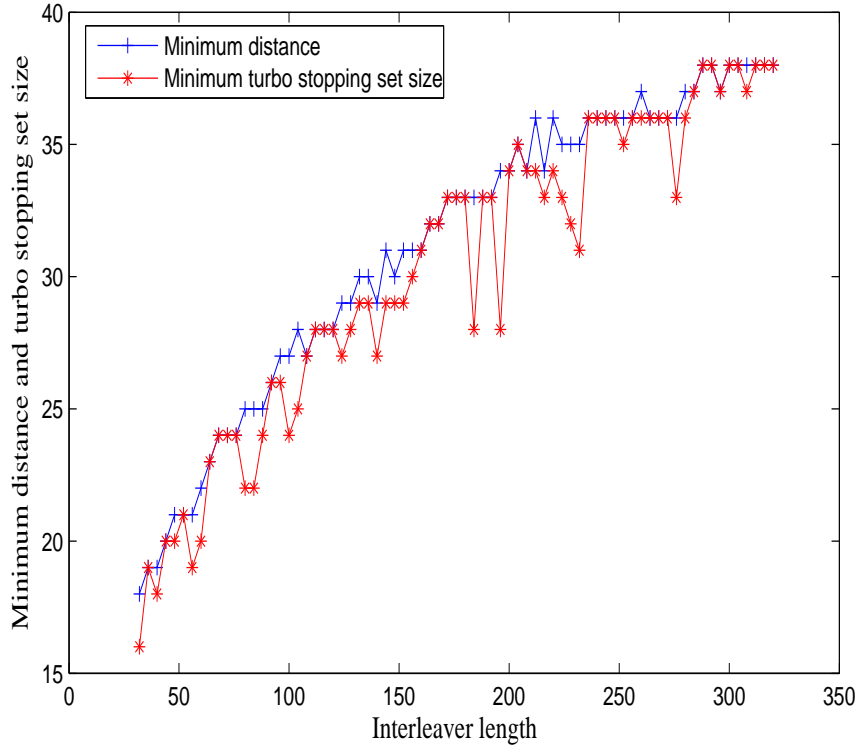}
\end{center}
\caption{\label{fig:dmin_data_matlab} {Minimum distance and minimum turbo stopping set size as a function of interleaver length for distance-optimized DRP interleavers with a dither length of $4$.}}
\end{figure}

We have applied the algorithm from Section~\ref{GPB} with the
improvements from \cite{ros03_12} on a few example codes. The example
codes are constructed without considering turbo stopping sets.
Consider the $(828,270,36)$ turbo code constructed by Crozier \emph{et
  al.} in \cite{cro04}. The code has an optimized minimum distance, is
based on a DRP interleaver, and is constructed from nominal
rate-$1/2$, $\nu=3$ constituent codes defined by the parity-check
matrix $\mathbf{H}(D) = (1+D+D^3\;1+D^2+D^3).$ The last polynomial
which is irreducible and primitive has been chosen as the parity
polynomial making the constituent encoders recursive.  For this code
there are $59\,(0)$, $58\, (58)$, and $283\, (283)$ stopping sets
(codewords) of size (weight) $33$, $36$, and $37$, respectively. The
minimum turbo stopping set size is smaller than the code's minimum
distance. This is typically what happens for both short and
moderate-length distance-optimized DRP interleavers. With the uniform
interleaver, however, there are
\begin{displaymath}
\begin{split}
& 0.08538\, (0.08538), 1.245\cdot10^{-7}\, (1.245\cdot10^{-7}), 0.01958\, (0.01958), 0.66860\, (0.66860), 1.91184\, (1.90691), \\
& 0.01171\, (0.00598), 0.27896\, (0.22497), 2.55047\, (2.44926), 3.90238\, (3.77258), 0.50298\, (0.28068), \\
& 2.48949\, (1.83803), 7.22456\, (6.12582), 8.24370\, (6.66759), 7.72356\, (4.50883)
\end{split}
\end{displaymath}
stopping sets (codewords) of size (weight) $9$, $10$, $12$, $13$,
$14$, $15$, $16$, $17$, $18$, $19$, $20$, $21$, $22$, and $23$,
respectively. Thus, small-size turbo stopping sets is not a problem
with the uniform interleaver for these parameters. As another example
consider an interleaver length of $1200$. In \cite{cro04}, Crozier
\emph{et al.} constructed an impressive DRP interleaver of length
$1200$ with a dither length of $8$. The corresponding turbo code has
length $3600$, dimension $1194$, and minimum distance $49$. For this
code we have found turbo stopping sets of size $47$.  Results for a
range of interleaver lengths are given in
Fig.~\ref{fig:dmin_data_matlab}. The turbo codes are constructed from
the same nominal rate-$1/2$ constituent codes as the $(828,270,36)$
code above. For each interleaver length between $32$ and $320$, in
which the dither length is a divisor, we have found the best (in terms
of turbo code minimum distance) DRP interleaver with a dither length
of $4$. Also, for the same interleavers we have found the minimum
turbo stopping set sizes. These results are plotted in
Fig.~\ref{fig:dmin_data_matlab}. Note that for several interleaver
lengths the minimum turbo stopping set size is smaller than the
minimum distance.

\section{Conclusion and Future Work} \label{sec:conclu}

In this work we have considered the finite-length analysis of turbo
decoding on the BEC. In the same way as iterative BP decoding of LDPC
codes is simpler on the BEC than on other channels, turbo decoding can
also be simplified on this channel. Based on this simplified turbo
decoding algorithm we have introduced turbo stopping sets by adapting
the concept of stopping sets from the theory of iterative BP decoding
of LDPC codes. These turbo stopping sets characterize turbo decoding
on the BEC, and an exact condition for decoding failure has been
established as follows. Apply turbo decoding until the transmitted
codeword has been recovered, or the decoder fails to progress further.
Then the set of erased positions that will remain when the decoder
stops is equal to the unique maximum-size turbo stopping set which is
also a subset of the set of erased positions. Furthermore, we have
presented some improvements of the basic turbo decoding algorithm on
the BEC. The proposed improved turbo decoding algorithm has
substantially better error performance as illustrated by simulation
examples. In the second part of the paper an expression for the turbo
stopping set size enumerating function under the uniform interleaver
assumption was derived. Also, an efficient enumeration algorithm of
small-size turbo stopping sets for a particular interleaver was given.
The solution is based on the algorithm proposed by Garello \emph{et
  al.}  in 2001 to compute an exhaustive list of \emph{all} low-weight
codewords in a turbo code. In fact, it turns out that every
trellis-based weight spectrum computation or estimation algorithm for
turbo codes can be adapted to the case of finding small-size turbo
stopping sets. In particular, the impulse methods by Berrou and Vaton,
Vila-Casado and Garello, and Crozier \emph{et al.} can be adapted in a
straightforward manner.

One interesting topic for future work is the design of interleavers in
which one considers both low-weight codewords and small-size turbo
stopping sets. Trellis-based interleaver design algorithms can in a
similar manner be adapted to this problem using the extended trellis
module.

Finally, we remark that the findings in this paper can be adapted in a
fairly straightforward manner to other turbo-like codes, e.g., RA
codes, serial concatenated convolutional codes, and product codes.

\end{document}